\RequirePackage{fix-cm}
\documentclass[twocolumn]{svjour_mod}          % twocolumn
\smartqed  % flush right qed marks, e.g. at end of proof
\usepackage{breakcites}
\usepackage{graphicx}
\usepackage{amsmath}
\usepackage{amsfonts}
\usepackage{amssymb}
\usepackage{xcolor}
\usepackage{caption} 
\usepackage{subcaption}
\def\BP{\mbox{BP}}
\def\bp{w}

\def\CRM{\mbox{CRM}}
\def\BeP{\mbox{BeP}}
\def\RBeP{\mbox{R-BeP}}
\def\IBP{\mbox{IBP}}
\def\RIBP{\mbox{R-IBP}}
\def\Pois{\mbox{Poisson}}
\def\iprob{\eta}
\def\Bern{\mbox{Bernoulli}}
\def\Beta{\mbox{Beta}}
\def\Normal{\mbox{Normal}}
\begin{document}

\title{Restricted Indian Buffet Processes}




\author{Finale Doshi-Velez   \and Sinead A. Williamson
}
\thanks{Finale Doshi-Velez is supported in part by DARPA grant LORELEI HR0011-15-2-0022. Sinead Williamson is supported in part by NSF grant 1447721.}

\institute{F. Doshi-Velez \at
              Harvard Paulson School\\
29 Oxford Street\\
Cambridge, MA 02138 \\
              Tel.: +1-617-496-0964\\
              \email{finale@seas.harvard.edu}           %  \\
           \and
           S. Williamson \at
           McCombs School of Business\\
           University of Texas at Austin\\
           2110 Speedway\\
           Austin, TX 78705\\
           Tel.: +1-512-471-3322\\
           \email{sinead.williamson@mccombs.utexas.edu}
}

\date{ }%Received: date / Accepted: date}

\maketitle

\begin{abstract}
  Latent feature models are a powerful tool for modeling data with
  globally-shared features.  Nonparametric exchangeable models such as
  the Indian Buffet Process offer modeling flexibility by letting the
  number of latent features be unbounded.  However, current models
  impose implicit distributions over the number of latent features per
  data point, and these implicit distributions may not match our
  knowledge about the data.  In this paper, we demonstrate how the
  Restricted Indian Buffet Process circumvents this restriction,
  allowing arbitrary distributions over the number of features in an
  observation.  We discuss several alternative constructions of the
  model and use the insights gained to develop Markov Chain Monte
  Carlo and variational methods for simulation and posterior
  inference.

\keywords{Bayesian nonparametrics \and Latent feature models \and Indian Buffet Process}
\end{abstract}

\section{Introduction}\label{sec:intro}

Generative models are a popular approach for identifying latent
structure in data.  For example, a musical piece may be naturally
modeled as a collection of notes, each with associated frequencies.  A
patient's health may be naturally modeled as a collection of diseases,
each with associated symptoms.  The text of a news article may be
naturally modeled as a collection of topics, each with associated
words.  In each of these cases, we posit that there exists a small set
of underlying features that are responsible for generating the
structure that we observe in the data.

When the number of these underlying features is unknown, Bayesian
nonparametric models such as the Indian Buffet Process
(IBP)~\cite{ibp_gg} provide an elegant generative modeling approach.
Specifically, the IBP posits that there are an infinite number of
potential underlying features, but only a finite number of features
underlie any particular observation.  The IBP has been the foundation
for a variety of modeling applications including choice
behavior~\cite{dilan-choice}, psychiatric
comorbitities~\cite{ibp_psych}, network models~\cite{kurt-links}, blind source
separation~\cite{knowles-ica},  image
modeling~\cite{zhou-2009}, and time-series models~\cite{bphmm}.

Under the IBP, the prior distribution over the number of features
underlying an observation is governed by a single parameter $\alpha$.
The number of features in an observation is expected \emph{a priori} to
be distributed as $\Pois( \alpha )$.  The two-parameter~\cite{ibp_gg}
and three-parameter~\cite{ibp_power_law} extensions of the IBP retain
this strong requirement for Poisson-distributed feature cardinality.
Other non-parametric latent variable models such as the infinite
gamma-Poisson process~\cite{igap} and the beta-negative Binomial
process~\cite{broderick2015combinatorial,zhou2011beta} also exhibit a
Poisson distribution over the number of non-zero features.  Even IBP
variants that posit various kinds of correlations between observations
\cite{gupta2013factorial,miller2008phylogenetic} or features
~\cite{doshi2009correlated} retain the Poisson property on the number
of features in each observation.  \cite{caron12} somewhat relaxes the
Poisson constraint; their model allows the number of features
underlying each observation to follow a mixture of Poissons.

However, there may be situations in which we do not desire
Poisson-distributed marginals.  For example, power law behaviors are
common in networks and natural language.  In medicine, the number of patients visiting a
clinic without severe illnesses may be much more than predicted by a
Poisson distribution.  When modeling articles, we may wish to preclude
the possibility of having no topics represented.  Image data may come
with labels, and the text of the label might provide strong clues
about the number of objects we can expect to see in the image.  In
other settings, we may know exactly the number of active features
associated with an observation.  For example, when modeling audio
recordings, the number of speakers in each recording might be known.
The IBP does not provide the flexibility to put an arbitrary prior
distribution on the number of latent features in an observation; even
with the mixture of Poissons allowed by~\cite{caron12} we are
constrained to overdispersed distributions with full support on the
non-negative integers.

In this article, we present and describe the Restricted Indian Buffet
Process (R-IBP), a recently developed model that allows an arbitrary
prior distribution to be placed over the number of features underlying
each observation. Unlike the model of~\cite{caron12}, this
distribution can have arbitrary support, or even be degenerate on a
single value. The R-IBP was originally presented in~\cite{ribp}; this
paper extends upon that exposition.  We present several alternative
constructions, new insights, and novel efficient inference techniques.

\section{Background: Completely random measures and Infinite Exchangeable Matrices}\label{sec:bg}

Many Bayesian nonparametric models, including the IBP, can be expressed in terms of completely
random measures (CRMs) \cite{crm}. A completely random measure $\mu$
is a random measure consisting of a collection of atoms
$\mu=\sum_{i}\pi_i\delta_{\theta_i}$\footnote{Technically, a CRM
  can also include a deterministic, non-atomic component; however we ignore this for simplicity.} on some space $(\Theta, \mathcal{A})$ such
that for any disjoint subsets $A_1,A_2\in\mathcal{A}, A_1\cap A_2 =
\emptyset$, the masses $\mu(A_1),\mu(A_2)$ assigned to those subsets
are independent. 

The atom sizes $\pi_i$ and locations $\theta_i$ and are governed by a
L\'{e}vy measure $\nu(d\pi, d\theta)$; different choices of the
L\'{e}vy measure yield different properties. For example, the L\'{e}vy
measure $\nu(d\pi,d\theta) = c \alpha \pi^{-1}(1-\pi)^{\alpha-1}d\pi
H(d\theta)$ describes the homogeneous beta process \cite{hjort}, whose
name reflects the fact that the atom sizes $\pi_i$ are equal in
distribution to the limit as $I\rightarrow \infty$ of
$\mbox{Beta}\left(\frac{c\alpha}{I},
c\left(1-\frac{\alpha}{I}\right)\right)$ random variables.  The
L\'{e}vy measure $\nu(d\pi, d\theta) = \gamma \pi^{-1}e^{-\lambda \pi}
d\pi H(d\theta)$ describes the gamma process, whose atom sizes
similarly correspond to the infinitesimal limit of a gamma
distribution.  We will write an arbitrary CRM with L\'{e}vy measure
$\nu(d\pi, d\theta)$ as $\CRM\left(\nu(d\pi,d\theta)\right)$.

CRMs can be used to construct distributions over
matrices with exchangeable rows and infinitely many columns.  To do so, we first define a
\textit{directing measure} $\mu:=\sum_{i=1}^\infty \pi_i \delta_{\theta_i}\sim \mbox{CRM}\left(\nu(d\pi,d\theta)\right)$ to be a CRM
with L\'{e}vy measure $\nu$.  We then let $\zeta_n :=\sum_{n=1}^\infty z_{ni}\delta_{\theta_i}
\stackrel{\mbox{\tiny{i.i.d.}}}{\sim} \mbox{CRM}(g(\mu)), n=1,2,\dots$ be a sequence of CRMs
whose L\'{e}vy measure is some functional $g(\mu)$ of this directing measure $\mu$. Then, following de Finetti, the
sequence $\zeta_1,\zeta_2,\dots$ is an infinitely exchangeable sequence
of measures.  If we consider only the atom sizes $z_{ni}$ of these
measures, then we can transform this sequence of exchangeable measures
into a sequence of exchangeable vectors $Z_n = (z_{n1},z_{n2},\dots)$.  Stacking these (infinitely
long) vectors results in a matrix $\mathbf{Z} = (Z_1,\dots, Z_n)$ with exchangeable rows.

One of the most commonly used models in this class is the
beta-Bernoulli process~\cite{ibp_bb}, which defines a distribution over infinitely
exchangeable binary matrices.  The directing measure $\mu$ is
distributed according to a beta process
\begin{equation*}
\BP(c,\alpha,H):=\CRM\left(c \alpha \pi^{-1}(1-\pi)^{\alpha-1}d\pi
H(d\theta)\right),
\end{equation*}
where $c,\alpha>0$ and $H$ is a probability measure on $\Theta$.
Conditioned on $\mu$, the $\zeta_n$ are distributed according to a
Bernoulli process 
\begin{equation*}
\BeP(\mu) := \CRM\left(\delta_1(d\pi)\mu(d\theta)\right).
\end{equation*}
The number of non-zero entries in each row of the resulting matrix will be finite, but random; marginally, this number will be distributed as $\Pois(\alpha)$.

Since the beta process and the Bernoulli process form a conjugate
pair, we can integrate out the directing beta process measure and work
directly with the exchangeable sequence of binary vectors. When $c=1$, the resulting exchangeable distribution is known as the Indian Buffet Process~\cite{ibp_gg}, and the predictive distribution can be described in terms of the following analogy: Let each column of our matrix correspond to a dish in an
infinitely-long buffet, and each row correspond to a customer. The
first customer selects a $\mbox{Poisson}(\alpha)$ number of dishes.
When the $n$th customer arrives at the buffet, there are a finite
number of previously sampled dishes and an infinite number of
unsampled dishes. He selects a dish that has previously been sampled
$m_i$ times with probability $m_i/n$, and selects a
$\mbox{Poisson}(\alpha/n)$ number of new dishes. For general $c\neq 1$, the corresponding exchangeable process is known as the two-parameter IBP \cite{ibp_gg,ibp_bb}; a related restaurant analogy is given in \cite{ibp_gg}.

The Indian Buffet Process can be used as the basis for a latent
feature model where both the number of latent features exhibited by a
given data point, and the total number of latent features, are
unknown. In this context, each row of the matrix corresponds to a data
point, and each column corresponds to a latent feature; a non-zero
entry indicates that a given data point exhibits a given feature.

Different choices of CRMs yield different properties in the resulting
matrix. For example, the three-parameter Indian Buffet Process
replaces the beta process directing measure with a stable-beta
process; the resulting random matrix exhibits power-law behavior in
the total number of features exhibited in $N$ rows
\cite{ibp_power_law}.  If we combine a gamma process directing measure
with a sequence of Poisson processes, we obtain the infinite
gamma-Poisson process \cite{igap}, a distribution over integer-valued
matrices. Other exchangeable matrices constructed in this manner
include the beta-negative binomial
process~\cite{zhou2011beta,broderick2015combinatorial} and the
gamma-exponential process~\cite{gammaexponential}.

\section{Exchangeable Binary Matrices with Arbitrary Marginals: The Restricted Indian Buffet Process}
\label{sec:RIBP}

The class of exchangeable matrices described in Section~\ref{sec:bg}
offers significant modeling flexibility.  One property, however,
cannot be avoided by judicious choice of CRM: the distribution over
the number of non-zero entries per row is always marginally Poisson.
This property is a direct consequence of the complete randomness of
the underlying random measures $\mu$ and $\zeta_n$.  To show this
property, we observe that, regardless of choice of directing measure,
there will be some probability $\pi_i$ that the column $i$ is
non-zero.  Each column $i$ is chosen independently, resulting in a
binomial distribution over the number non-zeros entries per row.  With
infinite columns, the binomial distribution converges to a Poisson
distribution.

We can also show, intuitively, how imposing an arbitrary distribution
over the number of non-zero entries must break the complete
randomness.  Suppose that we know that each row of our matrix has
exactly $J$ non-zero entries.  Next, suppose that we observe $J$
non-zero entries in the first $k$ columns.  We know that the remaining
(infinite) entries must be zero with probability one; the
probabilities of the entries in the disjoint sets of columns $1\dots
k$ and $(k+1)\dots$ are no longer independent.  Complete randomness
has been broken.

The Restricted Indian Buffet Process (R-IBP), first introduced in \cite{ribp}, is a distribution over exchangeable binary matrices with an arbitrary distribution over the number of non-zero entries per row. 
In the following sections, we describe several equivalent formulations for the R-IBP. While the focus is on restricted versions of the Indian Buffet Process, the ideas in this section can  be similarly
applied to build other matrices with arbitrary marginals, as we will describe in Section~\ref{sec:extensions}.

\subsection{Construction of the R-IBP via Restriction in the de Finetti Representation}\label{sec:deFinetti}

The R-IBP was originally constructed (in \cite{ribp}) by manipulating the beta-Bernoulli process representation of the IBP. Recall from Section~\ref{sec:bg} that we can represent the
IBP as a mixture of Bernoulli processes, directed by
a beta process:
\begin{equation}
\begin{split}
\mu:=\sum_i\pi_i\delta_{\theta_i} \sim& \BP(c,\alpha,H)\\
\zeta_n:= \sum_iz_{ni}\delta_{\theta_i} \stackrel{\mbox{\tiny{i.i.d.}}}{\sim}& \BeP(\mu)\\
Z_n :=& (z_{ni})_{i=1}^\infty.
\end{split}\label{eqn:beb1}
\end{equation}
Since we are not interested in the locations $\theta_i$ of the atoms, we will employ a slight misuse of notation and write Equation~\ref{eqn:beb1} as:
\begin{equation}
\begin{split}
\mu \sim& \BP(c,\alpha,H)\\
Z_n \stackrel{\mbox{\tiny{i.i.d.}}}{\sim}& \BeP(\mu).
\end{split}\label{eqn:beb2}
\end{equation}

We can modify this construction to give a restricted model where the number of non-zero entries per row is constrained to be some integer $J$, by replacing the Bernoulli process in Equation~\ref{eqn:beb2} with a \textit{restricted} Bernoulli process
\begin{equation}
\RBeP(Z_n;\mu, f = \delta_J) \propto \begin{cases} \BeP(Z_n;\mu) & \mbox{if } \sum_iz_{ni} = J\\
0 &\mbox{otherwise.}
\end{cases}\label{eqn:JBeP_cases}
\end{equation}
where the associated normalizing constant is proportional to the
probability that a random sample from a Bernoulli process has total
mass $J$. More concretely, this gives
\begin{equation}
\begin{split}
&\RBeP\left(Z_n;\mu, f = \delta_J\right)\\
 =& \frac{\prod_{i=1}^\infty \pi_i^{z_{ni}}(1-\pi_i)^{1-z_{ni}}\mathbb{I}(\sum_iz_{ni}=J)}{\sum_{z'\in\mathcal{Z}}\prod_i\pi_i^{z'_i}(1-\pi_i)^{1-z'_i}\mathbb{I}(\sum_iz'_i = J)},\label{eqn:JBeP}
\end{split}
\end{equation}
where $\mathcal{Z}$ is the support of $\BeP(\mu)$.

This restricted Bernoulli process is the random measure obtained by conditioning the Bernoulli process on its total sum; it can be seen as a nonparametric extension of the conditional Bernoulli distribution \cite{chen:2000}. Clearly it is no longer a completely random measure: disjoint subsets of $Z_n$ depend on each other via the total sum.

More generally, we may wish to have some arbitrary distribution $f$ on the number of non-zero entries per row. We can obtain this by creating an $f$-mixture of the distributions described by Equation~\ref{eqn:JBeP}, so that the probability of a vector $Z$ is given by
\begin{equation}
\RBeP(Z; \mu, f) \stackrel{d}{=} f\left(\sum_iZ_i\right)\RBeP\left(Z;\mu, \delta_{\sum_i Z_i} \right).
\label{eqn:RBeP}
\end{equation}

We can substitute these restricted Bernoulli processes (Equations~\ref{eqn:JBeP}, \ref{eqn:RBeP}) for the Bernoulli processes in Equation~\ref{eqn:beb2}, yielding the following Restricted Indian Buffet Process:
\begin{equation}
\begin{split}
\mu \sim& \BP(c, \alpha, H)\\
Z_n \stackrel{i.i.d.}{\sim}& \RBeP(\mu, f).
\end{split}\label{eqn:rIBP-restrict}
\end{equation}
Since the $Z_n$ are identically and independently distributed given $\mu$, de Finetti's theorem tells us the resulting matrix $\mathbf{Z}=(Z_n)_{n=1}^N$ has exchangeable rows.

We note that even if we choose $f(z) = \mbox{Poisson}(z;\alpha)$, we
do not recover the IBP.  The IBP has $\mbox{Poisson}(\alpha)$
marginals over the number of non-zero elements in each row; however,
conditioned on observing some elements in a row, the number of
non-zero entries in the remaining elements are distributed according
to a Poisson-binomial distribution.  Complete randomness requires that
distribution over the non-zero elements in some subset of columns does
not depend on the number of non-zero elements in a disjoint subset of
columns.  In contrast, an R-IBP with $f(z) = \mbox{Poisson}(z;\alpha)$
will retain $\mbox{Poisson}(\alpha)$ as the conditional distribution
over the total number of non-zero entries, even after some entries
have been observed.

\subsection{Construction via Subsets of an Exchangeable Sequence}\label{sec:ex_con}
In Section~\ref{sec:deFinetti}, we saw how the R-IBP can be
represented using the combination of a beta process directing measure
and a sequence of restricted Bernoulli processes parametrized by this
measure.  Sometimes it is more convenient to work solely in terms of the exchangeable matrix $\mathbf{Z}$ (which has a finite number of non-zero columns), integrating out the (infinite-dimensional) directing measure $\mu$. We can make use of the IBP predictive distribution to represent the R-IBP without representing the underlying beta process; however care must be taken to ensure the correct distribution.

We can generate an IBP-distributed sequence $\mathbf{Z}^* = (Z^*_1,Z^*_2,\dots)$ of vectors using the buffet-based predictive distribution described in Section~\ref{sec:bg}. Since this sequence is infinitely
exchangeable, its law is invariant to shuffling the order of any
finite subset \cite{Aldous:1983}. A direct consequence of this is that any infinite
sub-sequence $\mathbf{Z}$ of $\mathbf{Z}^*$ is again infinitely exchangeable. Thus, we can construct an $\RIBP(c,\alpha,f)$-distributed matrix
$\mathbf{Z}$ by sampling a sequence of vectors $\mathbf{Z}^*\sim\IBP(c,\alpha)$, and including each proposed vector
$Z_n^*$ into our matrix $\mathbf{Z}$ with probability
$f(\sum_iZ^*_{ni})$.

 We note that this is directly equivalent to the
restricted Bernoulli process method described in Section
\ref{sec:deFinetti}: if we integrate out the directing measure, a
sequence of Bernoulli process-distributed measures is described by the
IBP. However, an undesirable property of the IBP-based procedure is that, unlike the Bernoulli process-based procedure, one must retain the entire sequence $\mathbf{Z}^*$ (or at least, its sufficient statistics) to generate the next candidate for $\mathbf{Z}$. If we generate our proposed distributions based on the column counts of $\mathbf{Z}$ rather than $\mathbf{Z}^*$, the resulting matrix will not have the desired law - and in general will not even be exchangeable.

To demonstrate this lack of exchangeability, we will attempt to construct a $\RIBP(c=1,\alpha,f=\delta_1)$ matrix $\mathbf{Z}$ by generating candidate vectors for $Z_n$ based only on the counts of $Z_{1:n-1}$. As shown by \cite{Fortini:Ladelli:Regazzini:2000} and
\cite{Aldous:1983}, a sequence is infinitely exchangeable iff
\begin{equation*} (Z_{N+1},Z_{N+2})|(Z_1,\dots,Z_{N}) \stackrel{d}{=} (Z_{N+2},Z_{N+1})|(Z_1\dots Z_N).
\end{equation*}
It therefore suffices to check whether
$(Z_2,Z_3)|Z_1\stackrel{d}{=}(Z_3,Z_2)|Z_1$. Let $P$ be the law of the
IBP with parameters $1,\alpha$, and let $P^*$ be the law of the proposed variant. Since our restricting function $f=\delta_1$, trivially we have $P^*(Z_1 = (1,0,0,\dots))=1$. We will compare $P^*(Z_2=(1,0,0,\dots), Z_3 = (0,1,0,\dots))$ and $P^*(Z_2 = (0,1,0,\dots), Z_3 = (1,0,0,\dots))$.

Under the Indian Buffet Process, we have $P(Z_2 = (1,0,0,\dots)|Z_1 = (1,0,0,\dots)) =
\frac{1}{2}e^{-\alpha/2}$ and $P(Z_2 = (0,1,0,\dots)|Z_1 =
(1,0,0,\dots)) = \frac{\alpha}{4}e^{-\alpha/2}$; therefore if we
restrict $Z_2$ to these two cases, $P^*(Z_2 = (1,0,0,\dots)) =
\frac{2}{2+\alpha}$ and $P^*(Z_2 = (0,1,0,\dots)) =
\frac{\alpha}{2+\alpha}$.

Following a similar argument, 
\begin{equation*}
  \begin{split}
    &P^*(Z_3 = (0,1,0,\dots)|Z_1=(1,0,0,\dots),Z_2 = (1,0,0,\dots))\\
    &=
    \frac{\alpha}{\alpha+6}
  \end{split}
\end{equation*}
and
\begin{equation*}
  \begin{split}
    &P^*((Z_3 = (1,0,0,\dots)|Z_1=(1,0,0,\dots),Z_2 = (0,1,0,\dots))\\ &= \frac{3}{6+2\alpha}.
  \end{split}
\end{equation*}    
So, 
\begin{equation*}
\begin{split}
  &P^*(Z_2=(1,0,0,\dots), Z_3 = (0,1,0,\dots))\\
  &= \frac{2}{2+\alpha}\frac{\alpha}{\alpha + 6} = \frac{2\alpha}{\alpha^2 + 8\alpha+12}
  \end{split}
\end{equation*}
and
\begin{equation*}
\begin{split}
  &P^*(Z_2 = (0,1,0,\dots), Z_3 = (1,0,0,\dots)) \\
  &= \frac{\alpha}{2+\alpha}\frac{3}{6+2\alpha} = \frac{3\alpha}{2\alpha^2 + 10\alpha + 12}.
\end{split}
\end{equation*}

Clearly, $(Z_2,Z_3)|Z_1\stackrel{d}{\neq}(Z_3,Z_2)|Z_1$ under the
proposed construction, meaning the resulting sequence is not
exchangeable. In order to construct an exchangeable sequence via the
IBP, we must record the entire IBP-generated sequence and then select
an appropriate sub-sequence.

\subsection{Construction via Tilting the Bernoulli Process}\label{sec:tilting}

A tilted CRM $\mu^*$ is a random measure obtained by scaling the law
$P_\mu$ of a CRM $\mu$ on $(\Omega,\mathcal{A})$ by its total mass,
according to some function $h(\Omega)$ \cite{lau2013}, so that
\begin{equation}
P_{\mu^*}(A) := \frac{1}{\mathbb{E}[h(\mu(\Omega))]}\int_A h(\nu(\Omega))P_\mu(d\nu).\label{eqn:tilting}
\end{equation}
For example, if $h(x) = e^{-\gamma x}$, then $\mu^*$ is said to be
exponentially tilted. Exponentially tilting a CRM yields a different
CRM \cite{lau2013}; for example an exponentially tilted
$\alpha$-stable process is equal (in distribution) to a generalized
gamma process \cite{brix1999}. In general, however, a tilted CRM will
not be a completely random measure. For example, if $h(x)=x^{-q}$ for
some $q>0$, then $\mu^*$ is said to be polynomially tilted and is no
longer a CRM. Random measures constructed via polynomial tilting
include the Pitman-Yor process \cite{pitmanyor} (obtained by
polynomially tilting an $\alpha$-stable process) and the beta-gamma
process \cite{jamesbetagamma} (obtained by polynomially tilting a
gamma process).

In Equation~\ref{eqn:RBeP}, the probability of a vector $Z_n$ under the
restricted Bernoulli process is given by its probability under the
Bernoulli process, scaled by a function $f$ of the number of nonzero entries in $Z_n$ (or equivalently, the total mass of the corresponding
random measure $\zeta_n = \sum_iz_{ni}\delta_{\theta_i}$).  Thus the
restricted Bernoulli process can be described as a tilted Bernoulli
process\footnote{Arguably, the
  tilted Bernoulli process nomenclature is perhaps a better fit for the R-IBP, since
  for arbitrary $f$ the ``restricted Bernoulli process'' is in fact a
  mixture of restricted distributions. However, the tilting
  interpretation was not apparent when the models described in this
  paper were first introduced in \cite{ribp}, so we continue to use
  original term ``restricted'' for consistency.} with the tilting function $h(x)=f(x)$.  

\subsection{Construction via the Normalized Beta Prime Process and Invariance with respect to the Directing Measure}
\label{sec:invariance}

As shown in Equation~\ref{eqn:beb2}, the IBP can be written as a
sequence of Bernoulli processes with a beta process directing measure
$\mu$.  If only a finite number $N$ of rows $Z_n$ have
been observed, our uncertainty about $\mu$ is described by a beta
process with parameters
$c+N,\frac{c\alpha}{c+N}H+\frac{1}{c+N}\sum_{n=1}^N\zeta_n$.  As $N$
tends to infinity, this posterior will tend towards the uniquely
defined directing measure $\mu$.

In contrast, the beta process directing measure $\mu$ for the R-IBP
can \emph{never} be uniquely determined, even with infinitely many
observations.  To show this, we can re-construct the R-IBP in terms of
a beta-prime process~\cite{betaprime}. A beta-prime
process-distributed CRM $\tau:=\sum_i\bp_i\delta_{\theta_i}$ is
obtained by transforming the atoms $\pi_i$ of a beta
process-distributed CRM $\mu:=\sum_i \pi_i\delta_{\theta_i}$ according
to $$\bp_i:=\frac{\pi_i}{1-\pi_i}.$$ The de Finetti representation of
the R-IBP can now be written as
\begin{equation}
\begin{split}
\tau:=\sum_i\bp_i\delta_{\theta_i} \sim& \mbox{Beta-prime}(c,\alpha,H)\\
J_n\sim&f\\
P(Z_n|\tau,f) =& \frac{\prod_i \bp_i^{z_i}\mathbb{I}(\sum_iz_{ni}=J)}{\sum_{z'\in\mathcal{Z}}\prod_i \bp_i^{z'_i}\mathbb{I}(\sum_iz_{ni}=J)}.\label{eqn:JBeP_w}
\end{split}
\end{equation}
The law $P(Z_n|\tau,f)$ is invariant to rescaling the $\bp_i$ by some
constant $e^\beta$, that is,
\begin{equation*}
\RBeP(Z;\{\bp_i\}, J)\stackrel{d}{=}\RBeP(Z;\{e^{\beta}\bp_i\}, J)
\end{equation*}
for any $\beta\in \mathbb{R}$.  Rescaling the beta-prime process weights
$\bp_i$ by $e^\beta$ is equivalent to rescaling the atoms $\pi_i$ of
the corresponding beta process according to the nonlinear function
\begin{equation}
\pi_i' = \frac{\pi_i e^{\beta}}{\pi_i e^\beta + 1 - \pi_i}, 
\label{eqn:Esscher}
\end{equation}
which describes the Esscher transform of a Bernoulli random variable
\cite{esscher}.  Intuitively, this scale invariance occurs because the
R-IBP first chooses the \emph{number} of non-zero entries $J_n \sim f$
and then selects \emph{which} entries will be non-zeros.  Conditioned
on $J_n$, the absolute scale of the weights $\pi$ no longer matters; only their
relative sizes are important.  

The connection between the restricted IBP and the beta-prime process
makes it possible to remove extra degree of freedom present in the
beta-Bernoulli (or beta-prime-Bernoulli) construction by fixing the
scale through a normalized beta-prime process. While this is
theoretically appealing -- it leads to a unique directing measure for
each infinite sequence -- it offers little practical advantage, due to
the lack of a tractable representation for such a process.

\section{Extensions and Variations}
\label{sec:extensions}

In Section~\ref{sec:RIBP}, we focused on exchangeable models based on
the IBP.  However, the same ideas apply to exchangeable models based
on other completely random measures.  One can also relax the
exchangeability assumption to allow partial exchangeability, leading
to models appropriate for data with observation-specific covariates.

\subsection{Restricted Exchangeable Matrices based on Different Completely Random Measures}

In Section~\ref{sec:deFinetti} we showed that the Restricted IBP can
be constructed by starting from the beta-Bernoulli process
representation of the IBP, and replacing the Bernoulli process with a
restricted Bernoulli process. Rather than start from the
beta-Bernoulli process, we could pick any pair of CRMs to generate an exchangeable sequence $(\zeta_n)_{n=1}^N$,
provided we can parametrize the $\zeta_n$ using the directing measure
$\mu$, for example if $\mu$ and $\zeta_n$ form a conjugate pair
\cite{Orbanz:2009}:
\begin{equation}
\begin{split}
\mu\sim& \mbox{CRM}\left(\nu(d\pi,d\theta)\right)\\
\zeta_n \stackrel{\mbox{\tiny{i.i.d.}}}{\sim}& \mbox{CRM}(g(\mu)), n=1,2,\dots\label{eqn:genmat}
\end{split}
\end{equation}
If the support of the random measures $\zeta_n$ in
Equation~\ref{eqn:genmat} consists almost surely of measures with a
finite number of non-zero atoms, the resulting exchangeable sequence
can be interpreted as a row-exchangeable matrix with a finite  number of non-zero columns. Examples of such exchangeable
matrices include the beta-negative binomial process
\cite{zhou2011beta,broderick2015combinatorial} and the gamma-Poisson
process \cite{igap}. All such models exhibit the property that the
total number of non-zero elements of a row are (marginally)
Poisson-distributed, a direct consequence of the complete randomness
of the underlying random measures (as described in
Section~\ref{sec:RIBP}).

For any such exchangeable model, we can restrict the support of the
$\zeta_n$ to generate an exchangeable matrix with restricted
support. If $\zeta_n\sim \BeP(\mu)$, this amounts to placing some
distribution $f$ over the number of non-zero entries, as described in
Section~\ref{sec:deFinetti}. For other choices of CRM, however, the
support of the unrestricted measure will not be limited to binary
vectors, presenting a wider range of possible restrictions. We discuss
three possibilities below.

\paragraph{Restricting the number of non-zero entries per row}

If $\zeta_n$ is distributed according to a Bernoulli process, imposing
a distribution over the sum of a row is equivalent to imposing a
distribution over the number of non-zero entries. For more general
CRMs, these two cases are not the same. We first consider imposing a
function $f\left(\sum_k\mathbb{I}(z_k>0)\right)$ on the number of
non-zero entries. This yields a restricted CRM with law
\begin{equation}
  \begin{split}
    &\mbox{R-CRM}^{(1)}\left(Z;g(\mu), f\left(\sum_k\mathbb{I}(z_k>0)\right)\right) \\
    \propto& f\left(\sum_k\mathbb{I}(z_k>0)\right) \mbox{CRM}(Z;g(\mu))\label{eqn:gennonzero}
    \end{split}
\end{equation}
where $CRM(g(\mu))$ is the law of the corresponding unrestricted CRM. 

\paragraph{Restricting the sum of each row}
We can also impose a function $f(\sum_kz_k)$ on the total sum of each
row (or equivalently the total mass of each measure $\zeta_n$),
yielding
\begin{equation}
  \begin{split}
    &\mbox{R-CRM}^{(2)}\left(Z;g(\mu), f\left(\sum_kz_k\right)\right) \\
    \propto&  f\left(\sum_kz_k)\right) \mbox{CRM}(Z;g(\mu))\label{eqn:gensum}
  \end{split}
\end{equation}

A special case of the construction in Equation~\ref{eqn:gensum} is
obtained when the directing measure $\mu$ is distributed according to
a gamma process with parameter $\alpha H$ for some probability measure
$H$, and the $\zeta_n$ are distributed according to a Poisson process
with mean measure $\mu$ \cite{igap}. In this case, if we restrict the
total sum of each row following Equation~\ref{eqn:gensum}, the
distribution over the $\zeta_n$ is equivalent to that given by the
following Dirichlet process-multinomial model:
\begin{equation*}
\begin{split}
\rho \sim& \mbox{DP}(\alpha, H)\\
J_n \sim &f\\
\zeta_n \sim& \mbox{Mult}(\rho, J_n)
\end{split}
\end{equation*}

\paragraph{Restricting the sum of each row and the value of each element}
The examples in Equations~\ref{eqn:gennonzero} and \ref{eqn:gensum}
give a taste of the sort of distributions available under this
construction. We can also specify more complex restrictions. For
example, we could generate an exchangeable binary matrix with $J$
non-zero elements by taking letting the $\zeta_n$ be a CRM with
integer-valued atoms, and restricting both the number of non-zero
elements to be $J$ and the values of the non-zero elements to be
one. If the directing measure $\mu$ is distributed according to a
gamma process, and the $\zeta_n$ are distributed according to a
Poisson process, each row corresponds to sampling $J$ entries using
conditional Poisson sampling from a Dirichlet-distributed random
measure.

\subsection{Restricted Partially-Exchangeable Matrices}\label{sec:partexch}

In Section~\ref{sec:RIBP}, we assumed that our data points (or
equivalently, the rows of our matrix) are exchangeable. We can however
modify the R-IBP to yield a partially exchangeable model appropriate
for data with observation-specific covariates.  For example, each
observation might have an associated label indicating a group membership $m \in
\{1,\dots,M\}$, and each group could have group-specific restricting
distribution $f_m$, so that  
\begin{equation*}
\begin{split}
\mu \sim& \BP(c, \alpha, H)\\
Z_n \stackrel{i.i.d.}{\sim}& \RBeP(\mu, f_{m(n)}).
\end{split}
\end{equation*}
where $m(n)$ is covariate describing the group for observation
$n$. The resulting matrix would be partially
exchangeable in the sense that the distribution is invariant to
permuting rows belonging to the same group.

This model would be appropriate where we have observation-specific
information about the number of non-zero features. For example, we
might wish to construct a topic model with different distributions
over the number of topics depending on the type of document (novels
contain many topics, news articles contain fewer topics). Or, we might
have a feature extraction task with labeling information indicating
the expected number of features per observation -- for example, in
image modeling we might have descriptions or low-level labeling.

\section{Simulation from the R-IBP}\label{sec:inf_prior}
\label{sec:simulation}

In Section~\ref{sec:RIBP}, we presented several constructions for the
R-IBP. These constructions result in a variety of approaches for sampling
from the R-IBP prior. In this section, we discuss exact and
approximate approaches for sampling from the R-IBP.

\subsection{Sub-sampling from an Exchangeable Model}
\label{sec:rej_IBP}
In Section~\ref{sec:ex_con}, we showed that the R-IBP can be
constructed by subset selection of an IBP-distributed sequence of
binary vectors. This directly suggests a scheme for generating exact
from the prior. We generate a sequence $\mathbf{Z}^*=(Z_n^*)$
according to the Indian Buffet Process predictive
distribution:\footnote{Here we use the one-parameter version,
  i.e. $c=1$. We could easily substitute the two-parameter predictive
  distribution; see \cite{ibp_gg}.}

\begin{equation*}
\begin{split}
m_{ni} =& \sum_{j=1}^{n-1}z_{nj}^*\\
K^+_n =& \sum_k\mathbb{I}(m_{ni}>0)\\
z_{ni}^* \sim& \mbox{Bernoulli}(m_{ni}/n) \mbox{ for } i=1,\dots, K^+_n\\
\lambda_n \sim& \mbox{Poisson}(\alpha/n)\\
z_{nj}^* =& 1 \mbox{ for } j=K_n^++1,\dots, K_n^+ + \lambda
\end{split}
\end{equation*}
Then, we include each $Z_n^* = (z_{n1}^*, z_{n2}^*,\dots)$ in our
sequence $\mathbf{Z}$ with probability $P(Z_n^*\in \mathbf{Z}) =
f(\sum_i z_{ni}^*)$.  Importantly, while not all the generated binary
vectors $Z_n^*$ are included in $\mathbf{Z}$, they \textit{are} all
included in the counts $m_{ni}$, ensuring exchangeability is
maintained.  Rejection sampling in an exchangeable model produces
perfect samples from the R-IBP, but can suffer from a low acceptance rate.

\subsection{Approximate Sampling with a Conditionally Independent Model}
\label{sec:approx_prior}

An alternative approach, inspired by the construction in
Section~\ref{sec:deFinetti}, is to explicitly sample the directing
measure $\mu\sim\BP(c,\alpha,H)$ (or, alternatively, sample $\tau\sim
\mbox{Beta-prime}(c,\alpha,H)$), and use this to sample a sequence
$\mathbf{Z}^* =(Z_n^*)$ of binary vectors. Practically speaking, we
cannot represent the entire infinite-dimensional measure $\mu$ (or
$\tau$). However, we can work with finite-dimensional approximations
to $\mu$ to produce both approximate and exact samples.  We describe
approximate approaches in this section and an exact approach in
section~\ref{sec:exact_prior}.

We first need to produce a finite set of weights $\pi$ that well
approximate the infinite-dimensional measure $\mu$.  We consider two
options:
\begin{itemize}
\item \emph{Weak Limit} One approach is to use a finite vector of beta
  random variables that converges (in a weak limit sense) to the beta
  process \cite{zhou-2009}, approximating $\mu$ with a vector $\tilde{\pi}
  = (\tilde{\pi}_1,\dots,\tilde{\pi}_I)$, where
\begin{equation}
\tilde{\pi}_i \stackrel{\tiny{iid}}{\sim}\mbox{Beta}\left(\frac{c\alpha}{I},c-\frac{c\alpha}{I}\right).\label{eqn:weaklimit}
\end{equation}
\item \emph{Size-Ordered Stick-breaking Representation}
Another approach is to transform the arrival times of a unit-rate
Poisson process based on the beta process L\'{e}vy measure
\cite{rosinski,fergusonklass}. This approach gives exact samples from
the size-ordered atoms of the beta process. In the special case where
$c=1$, this yields a simple stick-breaking construction
\cite{ibp_stick}:
\begin{equation}
\begin{split}
u_i \sim& \mbox{Beta}(\alpha, 1)\\
\pi_i =& \prod_{j=1}^iu_i.\label{eqn:stick}
\end{split}
\end{equation}
If we let our truncated approximation $\tilde{\pi_i} = \pi_i$ for $i =
1 \dots I$ and $\tilde{\pi_i} = 0$ for $i > I$, we obtain an
approximation to a sample from a beta process.
\end{itemize}
Given an approximate sample $\tilde{\pi} = (\tilde{\pi}_1,\dots,
\tilde{\pi}_I)$ from our directing measure, there are a number of
methods to simulate $Z_n\sim \BeP(\tilde{\pi})$. We discuss several 
approaches below.

\subsubsection{Rejection Sampling}
\label{sec:rej_BeP}

\paragraph{Using a Bernoulli Process Proposal}
Conditioned on $\tilde{\pi}$, it is straightforward to sample binary
vectors $Z^*$ according to a Bernoulli process, by sampling $z^*_i
\sim \mbox{Bernoulli}(\pi_i), i=1,\dots,I$. We can use these binary
vectors as proposals in a rejection sampler. If $f$ is the desired
distribution over the number of non-zero entries per row, we accept a
proposal $Z^*$ with probability $f(\sum_iz^*_i)$.  Because we have
explicitly instantiated (an approximation to) the directing measure $\mu$, the rows of $Z$ are i.i.d. and we do not need to maintain
the sufficient statistics of the rejected binary vectors.

\paragraph{Using a tilted Bernoulli process proposal}
The rejection sampling procedure using a Bernoulli process proposal
will give low acceptance rates---and therefore high computational
cost---if the target distribution $f$ differs significantly from the
$\mbox{Poisson}(\alpha)$ distribution implied by the IBP. We can
improve the acceptance rate---and hence ameliorate the computational
costs---by exponentially tilting the Bernoulli process likelihood, as
described in Section~\ref{sec:invariance}.

If we tilt a Bernoulli process (or, equivalently, scale the beta
process-distributed directing measure according to
Equation~\ref{eqn:Esscher} and use the scaled directing measure as the base measure for a Bernoulli process), we change the distribution over the number of non-negative entries \cite{bn}. If we restrict the tilted Bernoulli process, however, the distribution is not affected by the tilting parameter $\beta$, i.e.
\begin{equation*}
  \begin{split}
&\RBeP( (\tilde{\pi}_1,\dots, \tilde{\pi}_I))\\ \stackrel{d}{=} \, &
\RBeP\left(
\left(\frac{e^\beta\tilde{\pi}_1,}{e^\beta\tilde{\pi}_1+1-\tilde{\pi}_1}
\dots,
\frac{e^\beta\tilde{\pi}_1,}{e^\beta\tilde{\pi}_I+1-\tilde{\pi}_I}\right)\right).
\end{split}
\end{equation*}
 We can maximize the likelihood of getting exactly $J$ non-negative
entries, by setting $\beta$ to be the unique solution to
\begin{equation*}
J=\sum_{i=1}^I \frac{e^\beta\tilde{\pi}_i}{e^\beta\tilde{\pi}_i+1-\tilde{\pi}_i}\, .
\end{equation*}
Thus, we can first sample the number of features $J_n$ on the $n$th row from $f$, Esscher transform the weights $\tilde{\pi}_i$ to maximize the chance of getting exactly $J_n$ non-zero entries, and then sample $Z_{n}$ using the transformed weights.  For computational
efficiency, the transformed weights can be cached for each
value of $J_n$. 

\paragraph{Discussion of approximation quality}
As $I\rightarrow \infty$, both the weak-limit approximation of
Equation~\ref{eqn:weaklimit} and the stick-breaking construction of
Equation~\ref{eqn:stick} will give exact samples from the R-IBP.
However, a finite $I$ will introduce errors.  When a stick-breaking
representation for $\mu$ is used, then we know that all weights
$\pi_j$, $j > I$ will be less than $\pi_I$.  In particular, the
iterative nature of the stick-breaking construction means that, if we
exclude the first $I$ atoms $\pi_1,\dots, \pi_I$, and scale the
remaining atoms by $\pi_I$, we are left with a (strictly ordered)
sample from the beta process.

We can consider the error introduced by this construction by considering the values of $z_{nj}$ that are excluded due to the truncation. If there are any non-zero elements $z_{nj}$
for $j > I$, our rejection probability will not be correct. Since the weights $\pi_j, j>I$ are described by a scaled beta process, we know that the number of excluded non-zero elements will be distributed as $\mbox{Poisson}(\alpha\pi_I)$. So, with probability $1-\mbox{Poisson}(0;\alpha \pi_I) =1 - \exp(-\pi_I\alpha)$ the true sum $\sum_i^\infty
z_{ni} \neq \sum_i^I z_{ni}$ and thus we may incorrectly reject or accept a proposal..

We will reject incorrectly if there are any non-zero elements $z_{nj}$
for $j > I$.  If we let $\mu = \sum_{i=1}^\infty \pi_i
\delta_{\theta_i}$, where the $\pi_i$ are strictly size-ordered, and
sample $Z_n^*$ from the Bernoulli process $\BeP(\mu)$, it follows that
the distribution over the number of non-zero elements for $z_{nj}, j >
I$ is given by $\sum_{i=I}^\infty z_{ni}^* \sim \mbox{Poisson}(\alpha
\pi_I)$.  The probability that all elements are zero is given by
$P(\sum_{i=I}^\infty z_{ni}^*=0) = \exp(-\pi_I\alpha)$.  Thus, with
probability $1 - \exp(-\pi_I\alpha)$, the true sum $\sum_i^\infty
z_{ni} \neq \sum_i^I z_{ni}$ and thus we may reject incorrectly.

Specifically, in the case where $f=\delta_J$, three possible outcomes
exist when we propose a binary vector $Z^*$ from a size-ordered
truncated approximation $\BeP((\pi_i,\dots, \pi_I))$:
\begin{enumerate}
\item $\sum_{i=1}^I z_{i}^* > J$: We reject the proposal. This is always correct. 
\item $\sum_{i=1}^I z_{i}^* = J$: We accept the proposal. However, if
  the truncated tail has $\sum_{i=I+1}^\infty z_{i}^*>0$, we should
  really have rejected.  Our decision is correct with probability
  $P(\sum_{i=I+1}^\infty z_i^* = 0) = \exp(-\pi_I \alpha)$.
\item $\sum_{i=1}^I z_{i}^* < J$: We reject the proposal.  However, if
  $\sum_{i=1}^{I}z^*_{i}=J-k$ but the truncated tail has
  $\sum_{i=I+1}^\infty z_{i}^*=k$, we will really should have
  accepted.  Our decision is correct with probability
  $1-P(\sum_{i=I+1}^\infty z_i^* = J-\sum_{i=1}^I z_{i}^* )=
  1-\mbox{Poisson}( J - \sum_{i=1}^I z^*_{i} ; \pi_I \alpha )$.
\end{enumerate}
We will use this enumeration to construct an exact sampler in
section~\ref{sec:exact_prior}.

\subsubsection{Sampling using Inclusion Probabilities}
\label{sec:inc_prior}

Even with tilting, rejection sampling can be computationally expensive
if $f$ differs significantly from $\mbox{Poisson}(\alpha)$. Given
a finite-dimensional approximation to $\tilde{\pi}$ to the directing
measure, an alternative is to use a draw-by-draw procedure based on
computing the inclusion probabilities $P(z_{ni}|\tilde{\pi},k_n)$ of each feature
\cite{aires}.\footnote{More generally, \cite{bh} lists over 50 ways to
  sample without replacement with unequal weights in the finite case.}
The marginal inclusion probability $\iprob_{k;J}$ that feature $k$ is
included in a sample of size $J$ is given by
\begin{equation}
  \iprob_{k;J} = \tilde{\pi}_k \frac{ S_{J-1}^{I-1}(
  \tilde{\pi}_1,...,\tilde{\pi}_{k-1},\tilde{\pi}_{k+1},...,\tilde{\pi}_I ) }{ S_J^{I}(
  \tilde{\pi}_1,...,\tilde{\pi}_I ) }\label{eqn:incprob}
\end{equation}
where $S_J^I$ corresponds to the probability of sampling $J$ elements
from the set of $I$ features if each feature was chosen independently with
probability $\tilde{\pi}_i$:
\begin{equation}
\textstyle S_J^I = \sum_{s \in A_J(I)} \prod_{k \in s} \tilde{\pi}_k \prod_{j \ni s} ( 1 - \tilde{\pi}_j )
\label{eqn:inclusion}
\end{equation}
where $A_J(I)$ is the set of all samples of size $J$ that can be drawn
from the elements $I$.  Fortunately, there is a recursion for
calculating the values $S_J^I$ can be computed in $O(I^2)$-time:
\begin{equation}
S_J^I = \tilde{\pi}_I S_{J-1}^{I-1}( \tilde{\pi}_1,...,\tilde{\pi}_{I-1} ) + ( 1 - \tilde{\pi}_I ) S_{J}^{I-1}( \tilde{\pi}_1,...,\tilde{\pi}_{I-1} )
\end{equation}
and thus with appropriate caching, all of the elements $\iprob_{k,J}$
can be computed (and cached) in $O(I^3)$ time, and any Esscher
transform (Equation~\ref{eqn:Esscher}) of the $\tilde{\pi}_k$ can be used in the
recursions above.

Given the marginal inclusion probabilities $\iprob_{ik}$, we now have a
draw-by-draw algorithm for sampling a row $Z_n$ from the prior:
\begin{enumerate}
\item Sample the total number of features $J_n \sim f$.
\item Set $J = J_n$.
\item For each feature $k \in \{1,\dots, I\}$:
  \begin{enumerate}
  \item Sample $z_{nk} \sim \Bern( \iprob_{k, J} )$.
  \item If $z_{nk} = 1$, then decrement $J \leftarrow J - 1$. 
  \end{enumerate}
\end{enumerate}

\paragraph{Discussion of approximation quality}
If a size-ordered stick-breaking representation is used to approximate
the weights $\pi$, then we can directly bound the errors on the
inclusion probabilities as functions of the truncation level $I$, the
size of the smallest instantiated weight $\pi_I$, and the function
$f$.  To do so, we first expand the expression for the probabilities
$S^\infty_J$, starting with equation~\ref{eqn:inclusion}:
\begin{eqnarray*}
  \textstyle S_J^\infty &=& \sum_{s \in A_J(I)} \prod_{k \in s} \pi_k \prod_{j \ni s} ( 1 - \pi_j )\\
  &&+ \sum_{s \ni A_J(I)} \prod_{k \in s} \pi_k \prod_{j \ni s} ( 1 - \pi_j ) \\ 
  &=& \exp(-\pi_I\alpha) \sum_{s \in A_J(I)} \prod_{k \in s} \pi_k \prod_{j \ni s, j \leq I} ( 1 - \pi_j )\\
  &&+ \sum_{s \ni A_J(I)} \prod_{k \in s} \pi_k \prod_{j \ni s} ( 1 - \pi_j ) \\
 &=& \exp(-\pi_I\alpha) S^I_J + \sum_{s \ni A_J(I)} \prod_{k \in s} \pi_k \prod_{j \ni s} ( 1 - \pi_j ) 
\end{eqnarray*}
where $A_J(I)$ are still the sets in which all $J$ non-zero entries
occur in the first $I$ columns.  The second line follows because the
probability of that all the columns $j > I$ are zero is $\exp(-\pi_I
\alpha)$.  

Since the probability of at least on non-zero element in columns $j >
I$ is $1 - \exp(-\pi_I\alpha)$, the second term is bounded between $0$
and $1 - \exp(-\pi_I\alpha)$.  Thus we can bound the inclusion
probabilities
\begin{eqnarray*}
  &&\iprob_{k;J}= \pi_k \frac{ S_{J-1}^{\infty}(
  \pi_1,...,\pi_{k-1},\pi_{k+1},...,\pi_I ) }{ S_J^{\infty}(
  \pi_1,...,\pi_I ) } \\
  &\geq& \pi_k \frac{ e^{-\pi_I\alpha}S_{J-1}^{I-1}( \pi_1,...,\pi_{k-1},\pi_{k+1},...,\pi_I ) }{  e^{-\pi_I\alpha}S_J^{I}(
  \pi_1,...,\pi_I ) + ( 1 - e^{-\pi_I\alpha} ) }\\
  &\leq& \pi_k \frac{ e^{-\pi_I\alpha}S_{J-1}^{I-1}( \pi_1,...,\pi_{k-1},\pi_{k+1},...,\pi_I ) + ( 1 - e^{-\pi_I\alpha}) }{  e^{-\pi_I\alpha}S_J^{I}(
  \pi_1,...,\pi_I ) }
\end{eqnarray*}
As expected, the quality of the approximation depends not only
truncation $I$ (and associated $\pi_I$) but also on the values
$S^I_J$.  If the probability of sampling $J$ elements from the first
$I$ is low, then the approximation will be poor because it is likely
that additional columns would have been required to sample $J$ elements.

\subsection{Exact Sampling in a Conditionally Independent Model}
\label{sec:exact_prior}

We consider rejection sampling in an R-IBP with restricting function $f=\delta_J$, where we
accept or reject proposals $Z^*\sim \BeP((\pi_i,\dots,
\pi_I))$.  In Section~\ref{sec:rej_BeP}, working with a truncated
version of $\mu$ obtained using a stick-breaking representation means
that some proposals are erroneously rejected or accepted, due to the absence or presence of non-zero elements below the truncation.  In
particular, there were two cases in which we could make mistakes: if
$\sum_i^I Z^*_{ni} < J$, we might reject incorrectly; if $\sum_i^I
Z^*_{ni} = J$, we might accept incorrectly.  To circumvent the
uncertainty in these outcomes, we use a dynamic truncation to obtain
exact samples from the R-IBP by using retrospective
sampling~\cite{papaspiliopoulos2008retrospective}.
\begin{itemize}
\item Sample an initial truncated directing measure $\pi_1\dots\pi_I$ according to the size-ordered stick breaking
  representation of the Beta process (Equation~\ref{eqn:stick}).
\item For $n=1,\dots, N$, repeat the following proposal step until we have accepted a row $Z_n$:
  \begin{itemize}
  \item Sample $z^*_{1} \dots z^*_{I} \sim \pi_1\dots\pi_I$, and compute the sum $K^* = \sum_{i=1}^I z^*_{i}$.  
\begin{itemize}
\item If $K^*>J$, reject $Z^*$.
\item If $K^*=J$, accept with probability $  \exp(-\pi_I \alpha)$.
\item If $K^*<J$,
\begin{itemize}
  \item Reject with probability $1- \mbox{Poisson}( J - \sum_i z_{ni} ; \pi_I \alpha )$. 
  \item Otherwise, expand the representation by sampling new $\pi_i , z^*_{i}$ for  $i = I + 1, I+2,\dots$ according to the stick breaking representation,  until $\sum z^*_n = J$. Accept the resulting $Z^*$, and update $I$ and $\pi$ to incorporate the new atoms.
\end{itemize} 
\end{itemize}
\end{itemize}
\end{itemize}
We can adapt this procedure to arbitrary restricting function $f$, by first sampling a row count $J_n\sim f$ for each row.
The growth of the truncation level $I$ will depend on $f$; if $J_n \sim
f$ is large then we may have to expand to very large truncation levels
$I$.  Specifically, starting with a truncation level too small may
result in many, many rejections before the truncation level is
sufficiently expanded.  However, the samples that we do accept will be
from the correct R-IBP prior.

\subsection{Empirical Comparison of Simulation Methods}
We empirically compared the simulation approaches from
Sections~\ref{sec:rej_IBP}, \ref{sec:rej_BeP}, \ref{sec:inc_prior},
and~\ref{sec:exact_prior} by measuring the number of rejections and
CPU time required to generate samples from the R-IBP with
concentration parameter $\alpha=5$ and restricting function
$f=\delta_J$ for $J = \{2, 5, 8\}$.  We generated 25 samples of 100
observations from each of the five approaches: exact collapsed
rejection sampling (Section~\ref{sec:rej_IBP}), approximate
uncollapsed rejection sampling and tilted approximate uncollapsed
rejection sampling (Section~\ref{sec:rej_BeP}), approximate inclusion
sampling (Section~\ref{sec:inc_prior}), and exact uncollapsed
rejection sampling (Section~\ref{sec:exact_prior}).

\begin{figure*}[ht]
  \centering
  \includegraphics[width=\textwidth]{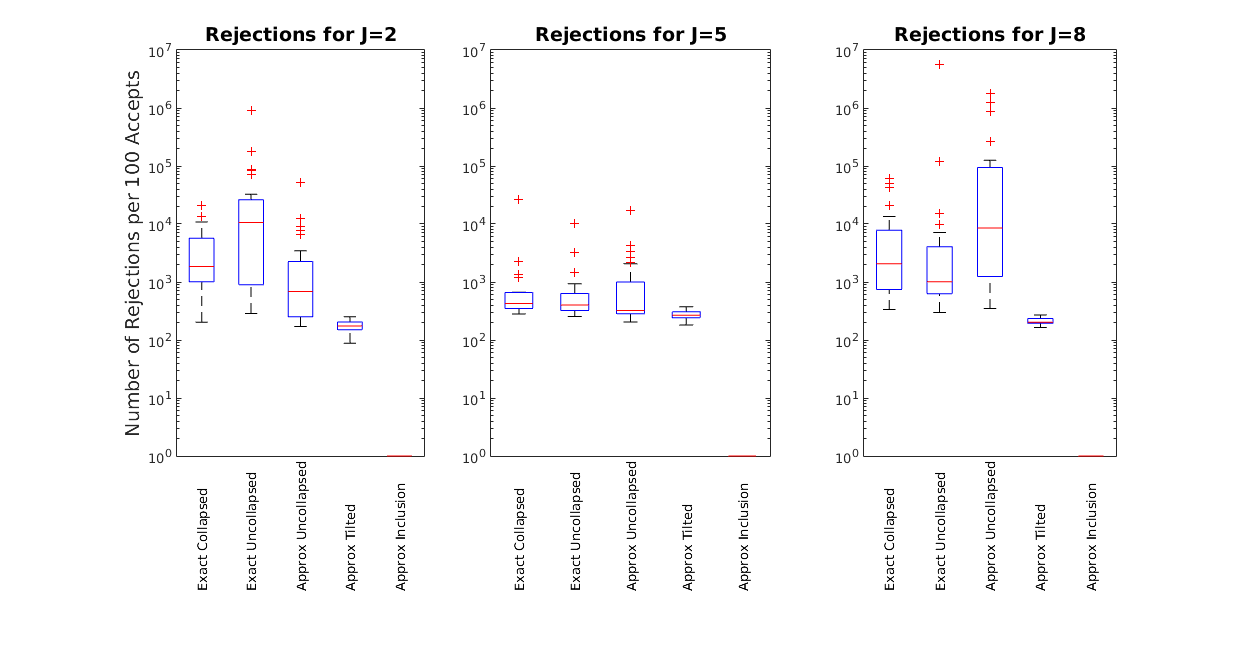}
  \caption{Rejections per 100 Acceptances}
  \label{fig:rej}
\end{figure*}

\begin{figure*}[ht]
  \centering
  \includegraphics[width=\textwidth]{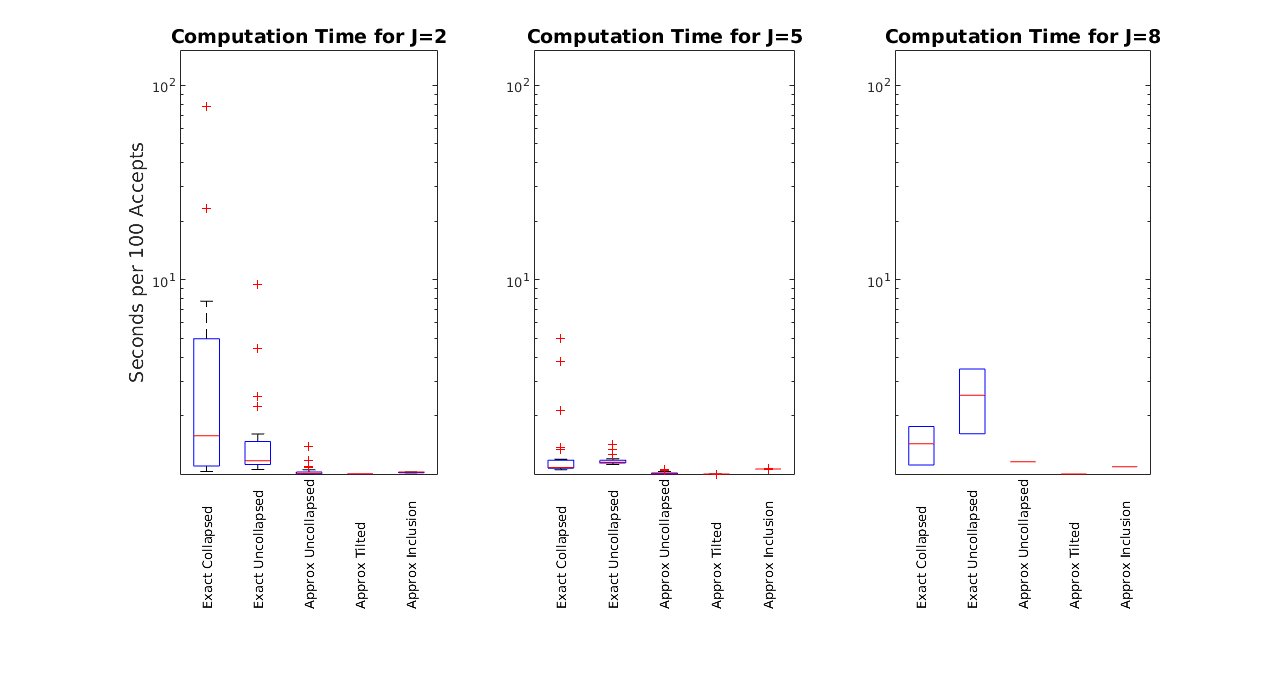}
  \caption{Time required for 100 Acceptances on a standard laptop}
  \label{fig:time}
\end{figure*}

Rejections per 100 observations are shown in figure~\ref{fig:rej}.  As
expected, rejection rates are lowest for $J=5$ because $\alpha=5$.
Inclusion sampling, a draw-by-draw procedure, has no rejections, and
tilting significantly reduces the number of rejections---and the
variance in the number of rejections---regardless of $J$.  The other
procedures all have large rejection rates varying over several orders
of magnitude.  Figure~\ref{fig:time} shows CPU time on a standard
laptop.  Again, the time to 100 acceptances is shortest when $J$ is
equal to the expected value of features $\alpha$.  The approximate
methods are faster than the exact methods, and the approximate tilted
rejection sampler is the fastest, closely followed by the approximate
sampler that uses inclusion probabilities.\footnote{The wall-clock
  time difference between the draw-by-draw procedure using inclusion
  probabilities and the approximate rejection samplers may be due in
  part due to Matlab vectorization; a draw-by-draw procedure requires
  a loop to sequentially compute whether a feature is present while
  the rejection sampler can sample all elements of $Z_n$ together.} 

\begin{figure*}[ht]
  \centering
  \includegraphics[width=\textwidth]{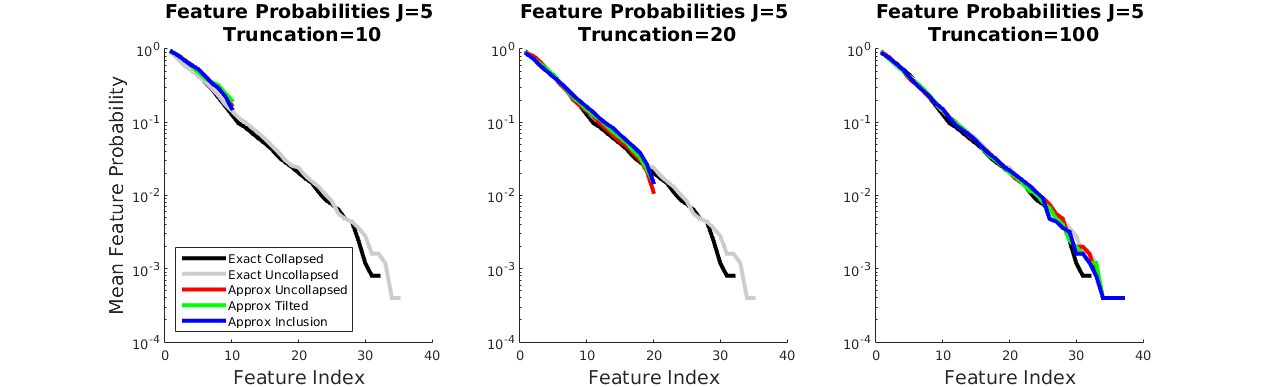}
  \caption{Mean of empirical feature probabilities, sorted in descending order for varying truncation levels and $J = 5$}.
  \label{fig:stick5}
\end{figure*}

\begin{figure*}[ht]
  \centering
  \includegraphics[width=\textwidth]{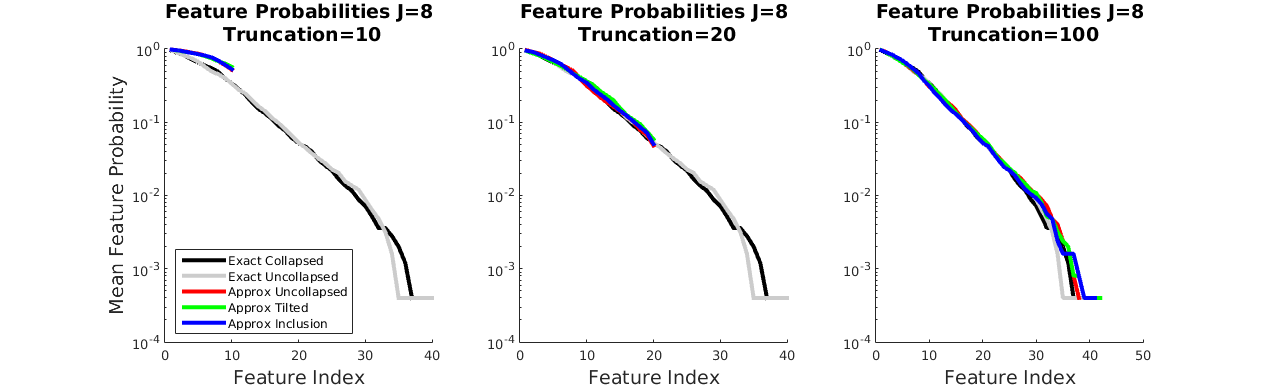}
    \caption{Mean of empirical feature probabilities, sorted in descending order for varying truncation levels and $J = 8$}.
  \label{fig:stick8}
\end{figure*}

Figures~\ref{fig:stick5} and~\ref{fig:stick8} show the mean of the
empirical feature probabilities, sorted in descending order, for
various truncation levels for $J=5$ and $J=8$.  When $J=5$, the exact
samplers instantiate between 30-40 hidden features.  The mean
probabilities of the approximate methods follow the exact probabilities
relatively closely even with truncations of $I=10$ or $I=20$, with
only slight over-estimation to account for the fewer features.  When
$J=8$, the exact methods tend to instantiate 35-45 features.  The
approximate methods have a noticeable over-estimation of feature
probabilities when the truncation $I$ is too small (e.g. $I=10$).
However, as the truncation is increased, the mean probabilities from
the approximate methods again closely match those from the exact
methods.  Interestingly, there do not seem to be large differences
between the different approximate methods.  These explorations suggest
that the approximate methods can be accurate,
computationally-efficient alternatives when the truncation is set to a
reasonable value.

\section{Posterior Inference in the R-IBP}
In this section, we present approaches for posterior inference in the
R-IBP. In Section~\ref{sec:inf_mcmc}, we present MCMC-based approaches
related to the simulation techniques described in
Section~\ref{sec:simulation}, and in Section~\ref{sec:inf_vi} we
present a computationally faster hybrid variational/MCMC approach for
posterior inference.

\subsection{MCMC-based Posterior Inference in the R-IBP}
\label{sec:inf_mcmc}

\subsubsection{Collapsed Inference using an Augmented Representation}\label{sec:collapsed_post}
In Section~\ref{sec:ex_con}, we showed that the R-IBP can be constructed by selecting subsets of an IBP, and in Section~\ref{sec:rej_IBP} we showed that this construction can be used to generate samples from the R-IBP prior. We can also use this construction to construct a collapsed Gibbs sampler, by reintroducing the discarded rows as auxiliary variables. For each data point $x_n$, let $Z_n$ be the associated latent R-IBP-distributed binary representation, let $t_n$ be an auxiliary variable indicating the number of discarded rows between observations $n-1$ and $n$, and let $c_n$ be an aligned auxiliary vector of counts associated with these discarded rows. Let $m_i = \sum_{n=1}^N(z_{ni} + c_{ni})$ be the total observed and auxiliary counts for the $i$th feature. 

\paragraph{Sampling $t_n$ and $c_n$}
When selecting a subset of the IBP, $t_n$ is the number of discarded samples between the $n-1$st and the $n$th accepted samples, and $c_n$ is the associated column counts. We can sample these directly, by sampling vectors $Z^*$ from the prior predictive distribution of the IBP given the remaining counts, $m+c_{-n}$. With probability $1-f(Z^*)$, we include $Z^*$ in the auxiliary counts $c_n$ and $t_n$, and sample another vector; with probability $f(Z^*)$ we do not include $Z^*$ in $c_n$ and stop our sampling procedure.

\paragraph{Sampling $Z_n$}

We have two options for sampling $Z_n$. We can propose an \emph{entirely new vector} $Z'$, by using the ultimate $Z^*$ obtained when sampling $c_n$ (i.e. the proposal $Z^*$ that we rejected from $c_n$) as a Metropolis Hastings proposal. Since the proposal is sampled from the prior predictive distribution of the R-IBP, we accept the proposal with probability
\begin{equation*}
\min\left(1,\frac{P(x_n|Z'_n, \Theta)}{P(x_n|Z_n, \Theta)}\right)
\end{equation*}

Alternatively, we can propose smaller changes to the current vector $Z_n$. For example, we could propose $z'_{ni} = 1-z_{ni}$, and accept with probability $\min(1, r)$ where
\begin{equation*}
\begin{split}
r=&\frac{P(x_n|z_{ni}', \mathbf{Z}_{- i},\Theta)P(z_{ni}'|m_{-z_{ni}}, N, \sum_nt_n)}{P(x_n|z_{ni}, \mathbf{Z}_{- i},\Theta)P(z_{ni}|m_{- z_{ni}}, N, \sum_nt_n)}\frac{q(z_{ni}'\rightarrow z_{ni})}{q(z_{ni} \rightarrow z_{ni}')}\\
=& \frac{P(x_n|z_{ni}',\mathbf{Z}_{- i},\Theta) m_{i, - z_{ni}}^{z'_{ni}}(1-m_{i,- z_{ni}}^{1-z'_{ni}} f(z'_{ni}+\sum_{k\neq i}z_{nk})}{P(x_n|z_{ni},\mathbf{Z}_{- i},\Theta)m_{i, - z_{ni}}^{z_{ni}}(1-m_{i,- z_{ni}}^{1-z_{ni}}f(z_{ni}+\sum_{k\neq i} z_{nk})}
\end{split}
\end{equation*}
Similarly, we could propose changing multiple entries at once (necessary if we have, for example, $f(x) = \delta_J(x)$), or adding/removing new features.

\subsubsection{Uncollapsed Inference with an Instantiated Latent Measure}\label{sec:uncollapsed_post}
If the distribution $f$ over the number of latent features per row differs significantly from that implied by the IBP,  the number of auxiliary features required in a collapsed scheme quickly becomes prohibitive. In practice, we found the computational cost of the sampler described in Section~\ref{sec:collapsed_post} was infeasible in most cases.

An alternative approach is to alternate sampling the latent measure
conditioned on the binary matrix, and vice versa, mirroring the
methods for sampling from the prior described in
Sections~\ref{sec:rej_BeP} and~\ref{sec:inc_prior}. Since we cannot
represent the entire measure $\mu$, we work with a finite-dimensional
approximation $\tilde{\pi} = (\tilde{\pi}_1,\dots, \tilde{\pi}_I)$, obtained either via a weak limit approximation or
via truncation in a stick-breaking
process. Sections~\ref{sec:inc_prior} and~\ref{sec:exact_prior} suggest
methods for bounding the resulting error, or using a dynamic
truncation to avoid such error.

\paragraph{Sampling $\mathbf{Z}|\tilde{\pi}$}

If the distribution $f$ is not degenerate on a single point, we can use the inclusion probabilities described in Section \ref{sec:inc_prior}, combined with $f$ and the data likelihood $P(X|Z,\Theta)$, to Gibbs sample the value of a single entry, using the conditional probabilities

\begin{equation}
\begin{split}
&P(z_{ni} = 1|\{\tilde{\pi}_1,\dots, \tilde{\pi}_I\}, \mathbf{Z}_{-ni}, X, \Theta)\\
  &\propto \tilde{\pi}_i \frac{S_0^{I-k_{n,-i}-1}(\{\tilde{\pi}_k:z_{nk}=0, k\neq i\})}{S_1^{I-k_{n,-i}}(\{\tilde{\pi}_k:z_{nk}=0 \mbox{ or }k=i\})}\\
  &\phantom{\propto} \cdot f(k_{n, -i}+1)P(X|z_{ni}=1, \mathbf{Z}_{-ni}, \Theta)\\[5pt]
&P(z_{ni} = 0|\{\tilde{\pi}_1,\dots, \tilde{\pi}_I\}, \mathbf{Z}_{-ni}, X, \Theta)\\ &\propto f(k_{n,-i})P(X|z_{ni}=0, \mathbf{Z}_{-ni}, \Theta),\label{eqn:gibbsZ1}
\end{split}
\end{equation}
where $k_{n,-k} = \sum_{j\neq k}z_{nj}$.

If the distribution $f$ \textit{is} degenerate on a single value $J$, we
cannot construct a Gibbs sampler that sequentially turns elements on
or off; doing so would change the number of features. Instead, we can
use the appropriate inclusion probabilities to sample the
location of each of the non-zero elements in a row, conditioned on the
other $J-1$ locations. Let $\ell_{nk}$ be the location of the $j$th non-zero entry. Then

\begin{equation}
\begin{split}
&P(\ell_{nj}=i|\{\tilde{\pi}_1,\dots, \tilde{\pi}_I\}, \ell_{-nj}, X, \Theta)\\
  & \propto \tilde{\pi}_i \frac{S_0^{I-k_{n,-i}-1}(\{\tilde{\pi}_k:z_{nk}=0, k\neq i\})}{S_1^{I-k_{n,-i}}(\{\tilde{\pi}_k:z_{nk}=0 \mbox{ or }k=i\})}\\
  &\phantom{\propto} \cdot f(k_{n, -i}+1)P(X|\ell_{nj}=i, \ell_{-nj}, \Theta)\label{eqn:gibbsZ2}
\end{split}
\end{equation}

The Gibbs sampling steps described in Equations~\ref{eqn:gibbsZ1} and
\ref{eqn:gibbsZ2} only change a single element of $\mathbf{Z}$ at a
time. This can lead to slow mixing. We can augment these Gibbs
sampling steps with Metropolis Hastings proposals generated from the
prior, using either the rejection sampling approach of
Section~\ref{sec:rej_BeP} or the inclusion probability approach of
Section~\ref{sec:inc_prior} to propose an entire row of the binary
matrix.

\paragraph{Sampling the latent measure}
Once we have sampled our binary matrix $\mathbf{Z}$, we must resample our latent feature weights $\tilde{\pi}$. Unfortunately, since the beta process is not conjugate to the restricted Bernoulli process, we cannot directly Gibbs sample $\tilde{\pi}$ given $\mathbf{Z}$. Instead, we use Metropolis-Hastings steps. Since the posterior distribution over $\tilde{\pi}$ given $\mathbf{Z}$ is likely to be similar to the poster distribution in the unrestricted IBP, we use the posterior distribution from the unrestricted IBP as a proposal distribution. The acceptance probability depends on the R-IBP likelihood (Equations~\ref{eqn:JBeP} and \ref{eqn:RBeP}).

\subsection{Hybrid Variational Inference in the R-IBP}
\label{sec:inf_vi}

The standard variational approach for the IBP~\cite{vibp} uses a
mean-field approximation which places independent distributions $q(z_{ni})$ over
each feature assignment $z_{ni}$.  Using such a factored distribution
is straightforward because each assignment $z_{ni}$ is drawn
independently given the weight $\pi_i$.  However, variational
inference in the R-IBP is challenging because fixing the number of active features $J_n$ introduces dependence between the $z_{ni}$, and because the implied prior
distributions over the marginal inclusion probabilities $\iprob_{ik}$ are complex. Further, the invariance of the likelihood to scaling the directing measure, as described in Section~\ref{sec:invariance}, can lead to inefficiencies in exploring the state space and computational difficulties due to very small atom sizes that may occur at certain scales.

We propose a hybrid variational for inference in the R-IBP that combines variational distributions over the feature assignments and model parameters with MCMC inference over the directing measure. As in Section~\ref{sec:uncollapsed_post}, we work with a finite dimensional approximation $\tilde{\pi}$ to the directing measure. We assume that the weights $\tilde{\pi}_i$ are
fixed during the variational update, and then alternate between
resampling the $\tilde{\pi}_i$ and updating the variational posterior on the
other variables.  We demonstrate this approach using a linear-Gaussian likelihood, where the data $X$ are assumed to be generated
by $ZA + \epsilon$, where $A$ is an $I\times D$ feature matrix with
independent normal priors $\Normal( 0 , \sigma_A^2 )$ on each value
and $\epsilon$ is a $N\times D$ matrix of independent noise drawn from
$\Normal( 0 , \sigma_n^2 )$. We note that the inference of the feature matrix $A$ is the same as in the standard IBP, and other likelihood models developed for the IBP can be substituted.

Specifically, the variables in the variational update are the feature
assignments $Z$, the feature values $A$, and the count of active
features per observation $J_n$.  We consider the following mean field
approximation for the variational inference:
\begin{itemize}
\item $q_{\phi_i}(A_i)$ independent Gaussian distributions with mean
  $\phi_i$, variance $\Phi_i$ on the posterior of the feature value
  vector $A_i$.
\item $q_{\nu_{ni}}(z_{ni})$ independent Bernoulli distributions,
  where $\nu_{ni}$ is the probability that $z_{ni}$ is active.
\item $q_{\gamma_{nk}}(J_n)$ multinomial distributions over the number
  of features in observation $n$, where $\gamma_{nk}$ is the
  probability that observation $n$ has $k$ active features.
\end{itemize}

Let ${W = \{\phi, \Phi, \nu, \gamma \}}$ be the set of variational
parameters, and let ${V = \{ A , Z , J_n \}}$ be the set of variables.
Because the actual and variational distributions belong to the
exponential family, coordinate ascent on the variational
parameters corresponds to setting the variational distribution ${\log(
q_{W_i} ) = E_{W_{-i}}[ \log( P( W , V | X , \Theta ) ) ]}$, where
$\Theta$ denotes the set of hyper-parameters~$\{ \sigma_n^2,
\sigma_a^2, \alpha, f \}$~\cite{wainwright}.

We focus on providing the variational updates for the parameters
associated with $Z$ and $J_n$, as the updates for the parameters
associated with $A$ (i.e., $\phi_k,\Phi_k$) are exactly the same as
in~\cite{vibp}.  The update for $\gamma_{nk}$ is:
\begin{equation}
\begin{split}
 \log( q_{\gamma_{n}}(J_n) ) =& E_Z[ \log P( J_n ) + \log P( Z_n | \tilde{\pi} , J_n ) ] \\ \nonumber
=& \textstyle\sum_{k=1}^K I(J_n = k)[ \log( f_{nk} ) + \nu_{ni} \log( \iprob_{ik} )\\
& + ( 1 - \nu_{ni} ) \log( 1 - \iprob_{ik} ) ]
\label{eqn:gamma_update}
\end{split}
\end{equation}
where $f_{nk}$ is the prior probability that observation $n$ has $k$
elements.  Exponentiating and normalizing, we recover the posterior
parameters $\gamma_{nk}$.

The update for variational parameters $\nu_{ni}$ for the assignments
$Z$ are also straightforward given the inclusion probabilities
$\iprob_{ik}$:
\begin{equation}
\begin{split}
\log( q_{\nu_{ni}}( z_{ni}) ) =& E_{J_n,Z_{-ni},A}[ \log( P( z_{ni} | Z_{-ni} , \tilde{\pi} , J_n ) ) \\
&+ \log( P( X_n | Z_n , A , \sigma_n^2 ) ) ] 
\label{eqn:nu_expectation}
\end{split}
\end{equation}
where the second term is again exactly the same as in~\cite{vibp}.
For the first term, we can write
\begin{equation}
\begin{split}
&E_{J_n,Z_{-ni}}[ \log( P( z_{ni} | Z_{-ni} , \tilde{\pi} , J_n ) ) ]\\
 =& E_{J_n,Z_{-ni}}[ z_{ni} I( J_n = k ) \log( \iprob_{ik} ) \\
&\qquad + ( 1 - z_{ni}) I( J_n = k ) \log( 1 - \iprob_{ik} ) ] \\
 =& \textstyle z_{ni} \sum_k \gamma_{nk} \log( \frac{ \iprob_{ik} }{ 1 - \iprob_{ik} } ) + c\, .
\label{eqn:nu_prior}
\end{split}
\end{equation}
Substituting equation~\ref{eqn:nu_prior} into
equation~\ref{eqn:nu_expectation}, we derive the update
\begin{equation}
\begin{split}
\textstyle \xi =& \textstyle \sum_k \gamma_{nk} \log( \frac{
  \iprob_{ik} }{ 1 - \iprob_{ik} } ) - \frac{1}{2\sigma_n^2} \bigg( -2
\phi_i X_n^T + Tr( \Phi_i )\\ &+ \phi_i \phi_i^T + 2 \phi_i ( \sum_{j
  \ne i } \nu_{nj} \phi_j^T ) \bigg) \\ \nu_{ni} =& \textstyle
\frac{1}{1 + \exp(-\xi) }\,.
\label{eqn:update_nu}
\end{split}
\end{equation}

The equations above show how to update the variational distributions
on $Z$, $A$, and $J_n$ given $\tilde{\pi}$.  During our inference process,
we iterate through the following steps:
\begin{enumerate}
\item Computing the partial variational posterior on $Z$, $A$, and $J_n$.
\item Sampling values of $Z$, $A$, and $J_n$ from the variational posterior.
\item  Sampling new values of $\tilde{\pi}$ given the sampled $Z$.  
\end{enumerate}
To resample $\tilde{\pi}$ given $Z$, we use the Metropolis-Hastings step
described in Section~\ref{sec:uncollapsed_post}, where we jointly
propose a new set of $\{\tilde{\pi}_1'...\tilde{\pi}_I'\}$ from the weak-limit
approximation to the IBP posterior distribution ${\tilde{\pi}_i' \sim \Beta(
  \frac{ \alpha }{I} + m_i , 1 + N - m_i )}$, where ${m_i = \sum_n
  z_{ni}}$.  Next we accept or reject using the beta process prior on
$\tilde{\pi}$ and the likelihood $P( Z | \tilde{\pi}' , J_n )$, which can be computed
using the inclusion probabilities $\iprob'_{ik}$.

\section{Evaluation}
We show a variety of evaluations to demonstrate the value of using the
R-IBP on real and synthetic data when we have some knowledge about the
marginals on the number of non-zero entries.

\subsection{Exploration with Synthetic Data}
\label{sec:toy_explore}

To explore the ability of the R-IBP to recover latent structure, we generated two datasets using a linear Gaussian model, and used the IBP, the R-IBP with an appropriate restricting distribution, and the partially-exchangeable R-IBP with labeling information described in Section~\ref{sec:partexch} to recover the latent structure.

\subsubsection{Knowledge about the Number of Latent Features Assists with Parameter Recovery.}  
One reason for using the R-IBP is when we have strong ideas of what a ``feature'' corresponds to, coupled with strong information about the number of such features. While an IBP may be able to model the data using a collection of features, these features may not correspond to our preconceived notions of features -- for example, the IBP might use multiple features where we expect a single feature.

To explore this, we generated a toy dataset with a total of 15 latent features. We generated 400 observations with 14 of the 15 latent features, and 100 observations with a single latent feature. We assumed a user-defined, observation-specific distribution over the number of features (corresponding to the partially exchangeable model described in Section~\ref{sec:partexch}). Specifically, if an observation $X_n$ contains $k_n$ features, we used a restricting distribution $f_n$ that is uniform over $k_n \pm 1$.

\begin{figure*}[ht]
  \centering
  \begin{subfigure}[b]{0.32\textwidth}
    \includegraphics[width=\textwidth]{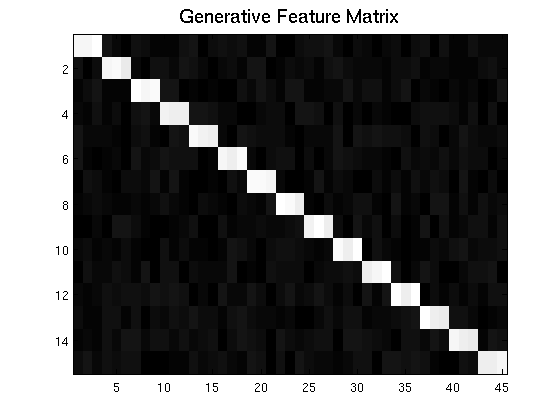}
    \caption{True A}
  \end{subfigure}
  \begin{subfigure}[b]{0.32\textwidth}
    \includegraphics[width=\textwidth]{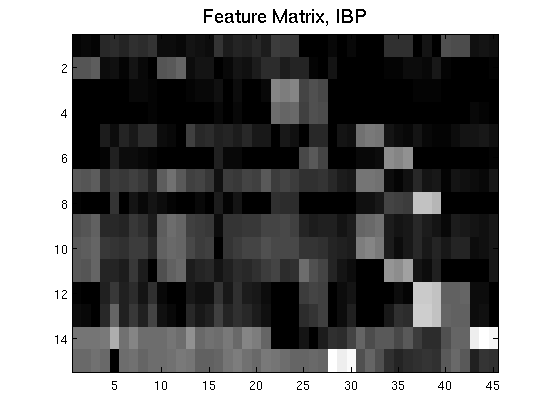}
    \caption{IBP A}
  \end{subfigure}
  \begin{subfigure}[b]{0.32\textwidth}
    \includegraphics[width=\textwidth]{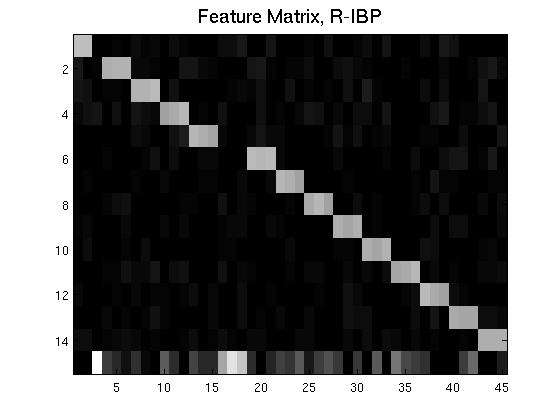}
    \caption{R-IBP A}
  \end{subfigure}

  \begin{subfigure}[b]{0.32\textwidth}
    \includegraphics[width=\textwidth]{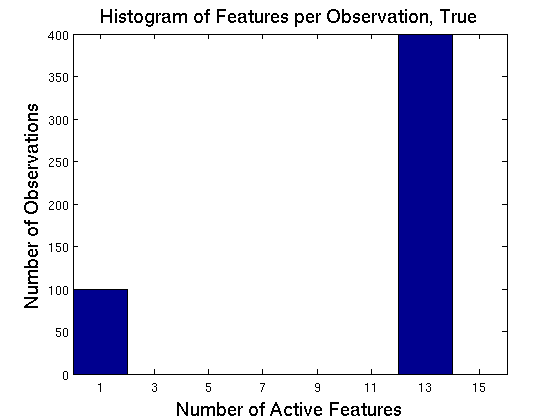}
    \caption{True Histogram over active features per observation}
  \end{subfigure}
  \begin{subfigure}[b]{0.32\textwidth}
    \includegraphics[width=\textwidth]{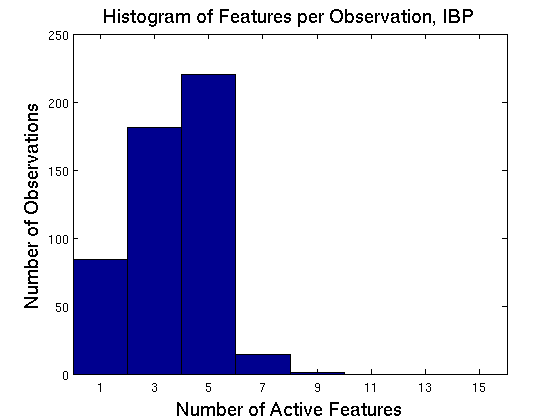}
    \caption{IBP Histogram over active features per observation}
  \end{subfigure}
  \begin{subfigure}[b]{0.32\textwidth}
    \includegraphics[width=\textwidth]{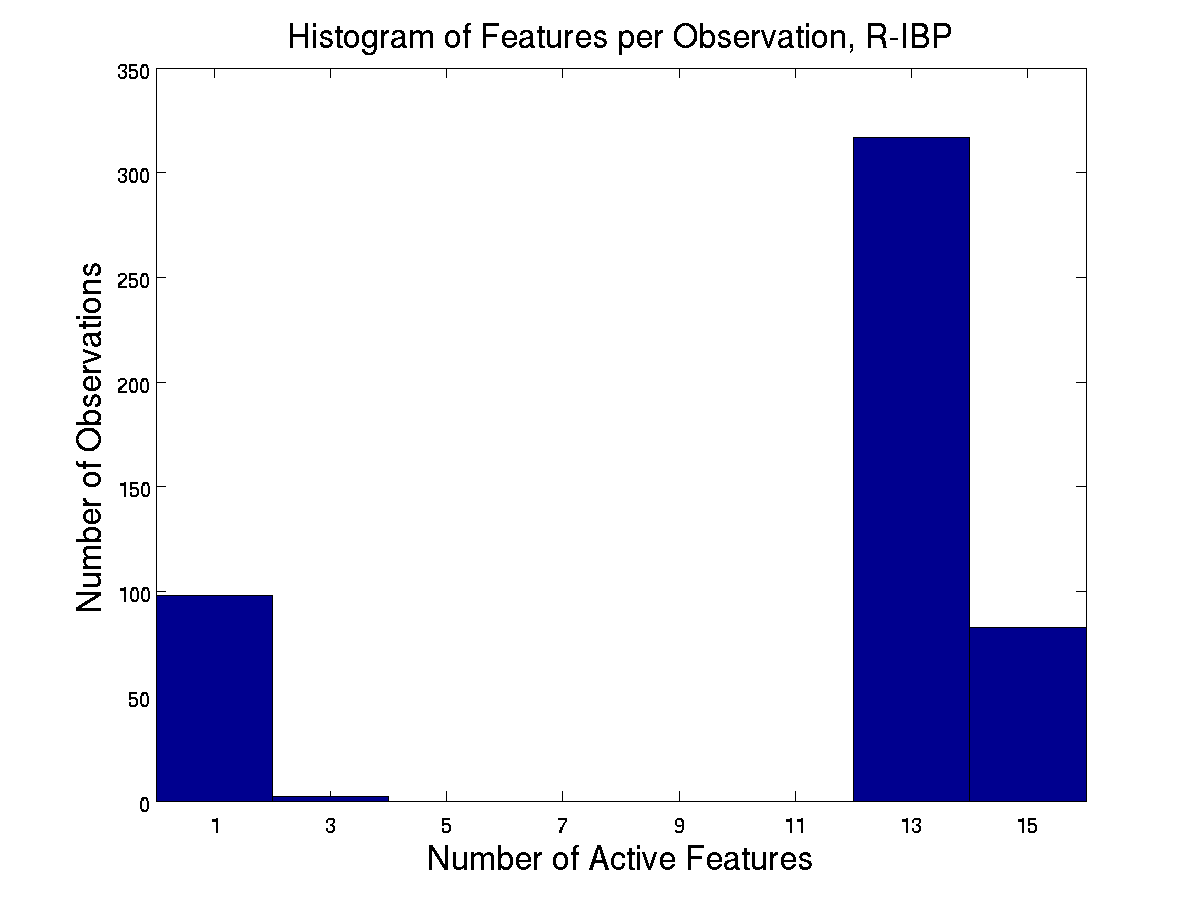}
    \caption{R-IBP Histogram over active features per observation}
  \end{subfigure}

\caption{Examples of features and counts of active features found by the
  variational inference for the R-IBP and the IBP on the toy data.
  The R-IBP recovers patterns much closer to the true features than the
  IBP, in which observations with just one feature tend to get assigned
  no features, while observations with many get a few generic features
  corresponding to most dimensions being active.  In contrast, the
  R-IBP recovers a histogram of features per observation that is much
  closer to the true distribution.}
\label{fig:toy-A}
\end{figure*}

Figure~\ref{fig:toy-A} shows qualitative results on the toy data.  The
first column shows the true features and the true distribution on the
number of active features in each observation.  Because many of the
features occur in many of the data sets, the IBP (center column) does not recover the
true features, nor does it recover a distribution of active features
that is close to the true distribution.  In contrast, the R-IBP (right
column) recovers a latent structure that is much closer to true
parameters.

\subsubsection{Knowledge about the Feature Distribution Assists with Predictive Performance}

While interpretable features are desirable, we do not want them to
come at the expense of predictive performance. To evaluate predictive
performance, we considered 500 observations from a one-inflated
Poisson model in which 80\% of the observations have one associated
latent feature and the remaining 20\% have a Poisson-distributed number
of associated latent features with mean $\lambda$.  Such a model might
be relevant when modeling patients in a typical clinical practice,
where most patients might have very simple complaints and a few
patients may have a very complex combination of diseases.  We apply
the Gibbs sampler for $\lambda = \{3, 6, 9, 12\}$; the concentration
parameter for the IBP was set to the mean number of features per
observation in each setting.

We explore two variants of the R-IBP: in the fully exchangeable
version, we know that observations come from a mixture distribution
but we do not know whether the observation is associated with the
spike or the slab; all observations have the same $f_n =
0.8\delta_{1}+0.2\mbox{Poisson}(\lambda)$.  In the partially
exchangeable version, we know to which mixture component the
observation belongs. If the observation belongs to the spike, we have
$f_n=\delta_1$, otherwise we have $f_n=\mbox{Poisson}(\lambda)$ This
assumption may be reasonable in many domains; for example, it may be
easy to tell if a patient has a simple or complex condition without
knowing explicitly what diseases a patient with complex diseases has.

\begin{figure*}
    \includegraphics[width=\textwidth]{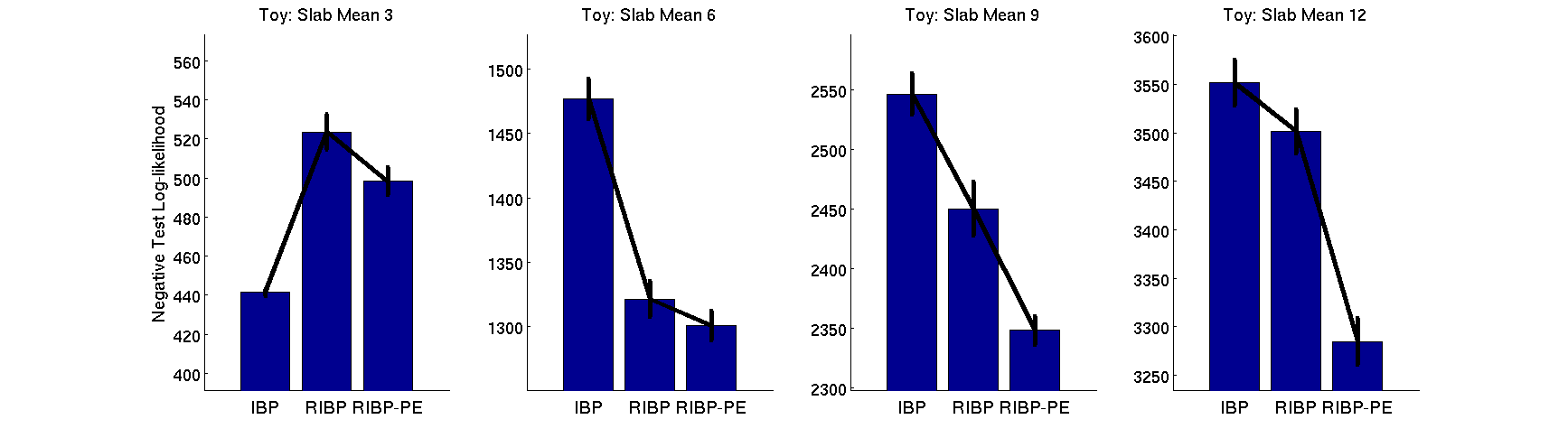}

\caption{Negative log-likelihoods (lower is better) for data from a
  one-inflated Poisson model with the mean of the Poisson $\lambda =
  \{3, 6, 9, 12\}$.  R-IBP is the fully exchangeable R-IBP model, whereas
R-IBP-PE is the partially-exchangeable R-IBP model where each
observation is associated with a covariate describing from which
distribution it comes.}
\label{fig:slab-explore}
\end{figure*}

We randomly held out 1\% of the data.  Figure~\ref{fig:slab-explore}
shows the negative log-likelihoods on the held-out data averaged over
5 runs of 500 iterations each (lower is better).  When the mean number
latent features in the slab distribution $\lambda = 3$, all
observations have few features, and the R-IBP variants performs
slightly worse than the IBP -- something we attribute to slower mixing
and therefore slower convergence, due to the lack of conjugacy.
However, as the slab mean $\lambda$ increases, the R-IBPs variants
consistently out-perform the IBP.  As expected, the
partially-exchangeable variant, in which each observation contains a
covariate describing whether it is a member of the spike or the slab,
does the best.

\subsection{Comparison on Multiple Real Data Sets}
We compare the two inference approaches for the R-IBP from
sections~\ref{sec:inf_mcmc} and~\ref{sec:inf_vi} to three IBP
baselines.  The hybrid variational IBP applies the same hybrid
variational approach to inference in the IBP as was developed for the
R-IBP in section~\ref{sec:inf_vi}.  We also compare to Gibbs
sampling in the IBP~\cite{ibp_gg} and the standard variational inference approach for the IBP~\cite{vibp}.  In all
cases, we the linear Gaussian likelihood model in which the data $X$
are assumed to be generated by $ZA + \epsilon$, where $A$ is an
$I\times D$ feature matrix with independent normal priors $\Normal( 0
, \sigma_A^2 )$ on each value and $\epsilon$ is a $N\times D$ matrix
of independent noise drawn from $\Normal( 0 , \sigma_n^2 )$.  Both the
Gibbs sampler and the variational methods were run for 300 iterations.
For the hybrid variational methods, the weights were resampled every
25 iterations of the coordinate ascent.  All methods were run 5 times.
A random 1\% of the data was held-out for evaluation.

We compare these methods on several data sets:
\begin{itemize}
\item The chord data set consists of a collection of three-note
  chords and single notes.  All 1320 three-note permutations and all
  12 single notes for the octave containing middle C were synthesized
  into wav files using MIDIUtil and FluidSynth; the power spectrum of
  these wav files was evaluated at every 10Hz between 0 and 1000Hz,
  resulting in a dataset with 100 dimensions. For the R-IBP, we used the partially exchangeable version, where we
  provided information stating that single notes had 1 latent feature
  in expectation $(k_n=1)$ and chords had 3 latent features in expectation ($k_n=3)$.
\item The newsgroup data-set is the subset of the 20 newsgroups data
  set from \textrm{http://www.cs.nyu.edu/~roweis/data.html},
  consisting of the counts for the top 100 words for 5000 documents.
  We arbitrarily set $k_n = \frac{L_n}{150}$, where $L_n$ was the
  length of the document.
\item The NPR data set consisted of the 365 features and the 365
  summaries from April 2013 to April 2014.\footnote{Source:
    http://www.npr.org/api/queryGenerator.php} The stories were
  processed through NLTK clean and we kept the 1964 most common
  words. We provided the information that the expected number of
  topics in a features story was $k_n = 1$ while the expected number
  of topics in a summary was $k_n = 5$.  
\end{itemize}
In all cases, the distribution $f$ was set to be uniform over $k_n \pm
1$.  We used a linear Gaussian likelihood in all cases.  

Likelihoods and training times for the toy problem and other problems
are shown in tables~\ref{tab:results1}, \ref{tab:results2} and
\ref{tab:results3}.  Here we see that the auxiliary information
provided by the R-IBP also translates into better likelihoods and not
just qualitatively better parameter recovery.  As expected, the
variational inference also runs significantly faster than the
MCMC-based approaches; however in some of the experiments the
variational approach yielded a lower quality estimate (shown most
clearly in the NPR dataset).

\begin{table*}[ht]
  \centering
\caption{Comparison of training set likelihoods for the R-IBP and the IBP.}
\begin{tabular}{|c||c|c|c|}\hline
\small
& Chord & Newsgroups & NPR \\ \hline \hline
Hybrid-Var. & -1.25e+05 & -2.33e+06 & -1.65e+07 \\
 R-IBP &  (-1.25e+05, -1.25e+05) & (-2.33e+06, -2.33e+06) & (-1.66e+07, -1.65e+07)  \\ \hline
Hybrid-Var. & -2.13e+05& -2.38e+06 & -6.49e+06  \\
 IBP&  (-2.13e+05, -2.13e+05) &  (-2.38e+06, -2.38e+06)  &(-6.50e+06, -6.48e+06)\\ \hline
Variational  & -2.13e+05  & -2.39e+06& -6.90e+06 \\
 IBP &(-2.13e+05, -2.13e+05) &  (-2.39e+06,  -2.39e+06) & (-6.96e+06,  -6.83e+06) \\ \hline
Gibbs R-IBP  & -1.33e+05  & -2.34e+06 & -4.89e+06  \\
&(-1.33e+05, -1.33e+05) & (-2.34e+06, -2.34e+06) & (-4.90e+06, -4.89e+06)\\ \hline
Gibbs IBP  & -1.25e+05    & -2.34e+06   & -5.15e+06  \\
& (-1.25e+05, -1.25e+05) & (-2.34e+06, -2.34e+06) & (-5.16e+06, -5.14e+06) \\ \hline
\end{tabular}
\label{tab:results1}
\end{table*}

\begin{table*}[ht]
  \centering
\caption{Comparison of test set likelihoods for the R-IBP and the IBP.}
\begin{tabular}{|l||c|c|c|}\hline 
\small
& Chord & Newsgroups & NPR \\ \hline \hline
Hybrid-Var.  & -2.11e+03& -2.38e+04 & -1.77e+05  \\
R-IBP&  (-2.13e+03, -2.09e+03) &  (-2.38e+04, -2.37e+04)  & (-1.80e+05, -1.74e+05) \\ \hline
Hybrid-Var. & -2.12e+03   & -2.43e+04  & -7.07e+04 \\
 IBP & (-2.13e+03, -2.11e+03) & (-2.43e+04, -2.42e+04) & (-7.14e+04, -7.00e+04) \\ \hline
Variational & -2.12e+03 & -2.43e+04 & -7.28e+04 \\
IBP&(-2.13e+03,  -2.11e+03) & (-2.43e+04, -2.42e+04) &(-7.33e+04, -7.22e+04)\\ \hline
Gibbs R-IBP  & -2.26e+03  & -2.37e+04  & -5.46e+04 \\
& (-2.29e+03, -2.24e+03) & (-2.37e+04, -2.37e+04) & (-5.48e+04, -5.44e+04) \\ \hline
Gibbs IBP  & -2.32e+03  & -2.37e+04  & -5.79e+04 \\
& (-2.34e+03, -2.29e+03)  &  (-2.38e+04, -2.37e+04) & (-5.81e+04, -5.77e+04) \\ \hline
\end{tabular}
\label{tab:results2}
\end{table*}

\begin{table*}[ht]
  \centering
\caption{Comparison of running times (in seconds) for the R-IBP and the IBP.}
\begin{tabular}{|l||c|c|c|}\hline
  \small
  & Chord & Newsgroups & NPR \\ \hline
  Hybrid-Var. & 1.43e+03 & 1.56e+05  & 1.02e+04 \\
  R-IBP &(1.42e+03, 1.44e+03) & (1.55e+05, 1.58e+05) &  (1.01e+04, 1.02e+04) \\ \hline
  Hybrid-Var.  & 9.68e+02  & 3.21e+04   & 1.65e+04 \\
  IBP & (9.60e+02, 9.76e+02) & (3.19e+04, 3.23e+04) & (1.63e+04, 1.66e+04)\\ \hline
  Variational & 1.05e+03  & 3.69e+04  & 1.50e+04 \\
  IBP &  (1.04e+03, 1.06e+03) & (3.67e+04, 3.72e+04) & (1.49e+04, 1.50e+04) \\ \hline
  Gibbs R-IBP & 3.56e+03  & 2.04e+04 & 9.97e+03\\
  & (3.52e+03, 3.60e+03) & (1.99e+04, 2.08e+04) &  (9.70e+03, 1.02e+04) \\ \hline
  Gibbs IBP  & 2.01e+03& 1.33e+04 & 7.39e+03 \\
   &  (1.99e+03, 2.02e+03) & (1.31e+04, 1.35e+04) & (7.20e+03, 7.58e+03) \\\hline

\end{tabular}
\label{tab:results3}
\end{table*}

\section{Discussion and Future Work}
The Restricted Indian Buffet Process is a useful tool for latent
feature modeling with a non-Poissonian number of latent features per
data point. In this article, we have expanded on the original
exposition \cite{ribp} by providing new representations that connect
the R-IBP to tilted CRMs and the scaled beta-prime process.  We also
provide several alternatives for exact and approximate simulation from
the R-IBP, as well as new inference algorithms, including a computationally efficient variational/MCMC hybrid algorithm.  

While the IBP often has reasonable performance on data sets with
arbitrary distributions over the number of features---rather than a
Poisson distribution---we find that additional knowledge about the
number of features can be very helpful if it is available.  In
particular, a common challenge when performing inference with the IBP
is that it often learns combinations of features as a single feature,
especially when there are correlations between features.  While these
feature combinations may reasonably represent the data, a latent
variable model that learns such grouped features will do poorly if
asked to make predictions on observations where that correlation is
not present.  With the R-IBP, it is possible to specify the expected
number of features in an observation, allowing us to discover features
with both better interpretability and generalization.  

In general, we see the most pronounced differences in situations where
we had strong prior knowledge about the number of features in a
dataset---such as the chord and toy examples.  Differences were less
pronounced in data sets such as newsgroups, where we made somewhat
arbitrary decisions about the potential number of features based on
document lengths; in general the IBP is a sufficiently flexible prior to capture posteriors with relatively small deviations from Poisson-distributions on the number of latent features, and in this case we actively decreased this flexibility. An interesting direction for further research would
be try to leverage less strong prior information---such as the
information in the NPR data set where some stories are features and
some stories are collections of multiple news summaries.  

More broadly, while we have focused on the Indian Buffet Process, the
concepts described in this paper are applicable to other nonparametric
models such as the beta-negative Binomial process or gamma-Poisson
process.  As we discussed in Section~\ref{sec:extensions}, the variety
of possible restrictions is much broader when considering non-binary
matrices, which are often used for modeling count data.  It will be
interesting to explore where restricted models can be effectively used
in this context; in principle different restrictions can allow domain
experts to encode a rich number of kinds of prior knowledge.

Finally, there is much to be explored on approaches for incorporating
the kinds of observation-specific restrictions described in this work.
The R-IBP has a natural interpretation as an IBP with arbitrary
distributions on the number of features in each observation.  However,
as we discussed in Section~\ref{sec:invariance}, there is an extra
degree of freedom when we specify the Restricted IBP with a beta
process or a beta-prime process.  Intuitively, this invariance arises
because conditioned on the number of latent features in an
observation, the scale of the weights no longer matters.  Any
restriction that can be viewed as conditioning will result in this
property.  In theory, working with a normalized beta-prime process
would remove this invariance; in practice, working with a normalized
beta-prime process is intractable.

However, there do exist other tractable normalized random measures
\cite{nrmi} such as the Dirichlet process and other and nonparametric
probability measures such as the Pitman-Yor process \cite{pitmanyor}.
These measures could be substituted for the beta-prime process in
Equation~\ref{eqn:JBeP_w}.  The resulting model could no longer be
interpreted as a restricted version of the IBP, but it is nonetheless
a valid model that may have very similar properties.  Having a more
potentially more tractable directing measure may assist in developing
robust and scalable inference techniques for restricted models.  

\begin{acknowledgements}
  The authors would like to thank Ryan P. Adams for numerous helpful discussions and suggestions, and Jeff Miller for suggesting the link to tilted random measures.
  \end{acknowledgements}
\bibliographystyle{apalike}%SandC/spbasic}      % basic style, author-year citations
\bibliography{ribp}   % name your BibTeX data base

\end{document}